\documentclass{aa}

\usepackage{graphicx}
\usepackage{txfonts}
\usepackage{hyperref}

\begin{document}

\title{Why is there cold gas inside the Local Bubble?}

\author{T.-E. Rathjen\inst{1,}\inst{2}\corrauth{rathjen@ph1.uni-koeln.de}
\and J. L. Linsky\inst{3}\email{jlinsky@jila.colorado.edu}}

\institute{Institute for Astrophysics, University of Cologne, Z\"ulpicher Str. 77, 50937 Cologne, Germany
\and Cluster of Excellence `Our Dynamic Universe' (DYNAVERSE)
\and JILA, University of Colorado and NIST, Boulder, CO 80309-0440, USA}

\date{}

\abstract
  {The Local Bubble (LB) extends approximately 100--200~pc from the Sun and contains $10^6$~K plasma and warm photoionised gas.
  Yet the Local Leo Cold Cloud (LLCC) and the Local Ribbon of Cold Clouds lie only 11--24~pc away, well inside the LB, at temperatures near 20~K.
  How such cold material forms and survives in this hotter environment has remained uncertain, and explanations based on colliding warm clouds are difficult to test observationally.}
  {We investigate whether LLCC-like gas can assemble locally in a feedback-driven environment or must be imported as pre-existing cold material, and whether it can survive there for several Myr.}
  {We used magnetohydrodynamic simulations of the interstellar medium from the \textsc{SILCC} Project that include self-consistent star formation, stellar-wind and supernova feedback, extreme-ultraviolet radiation, a spatially and temporally varying far-ultraviolet field, non-equilibrium chemistry, and cosmic-ray transport.
  We identified cold diffuse gas with $T<100$~K and $0<n_\mathrm{H_2}\leq2$~cm$^{-3}$, characterised its morphology and pressure environment, and used Lagrangian tracer particles to reconstruct its assembly, cooling, survival, and dispersal.}
  {The simulations produce cold diffuse gas in warm--hot, feedback-driven surroundings, where non-thermal pressure support is crucial.
  Across cold-gas interfaces, cosmic-ray pressure remains nearly continuous and establishes a pervasive pressure floor, while magnetic pressure partly offsets the external thermal-pressure excess.
  The structures persist in the cold diffuse phase for a median of 2.5~Myr, and most of their material remains cold even after they lose spatial coherence.
  The gas can assemble locally from nearby cold and thermally unstable warm material, developing filamentary or sheet-like morphologies.
  Its evolution to $\sim20$~K is driven not by adiabatic expansion but by non-adiabatic cooling.}
  {The existence of the LLCC inside the LB therefore does not require it to have entered as an already-formed cold cloud.
  LLCC-like gas can instead assemble locally and cool non-adiabatically in situ.}


\maketitle
\nolinenumbers

\section{Introduction}

Cold, dense gas is commonly associated with the swept-up shells and outer boundaries of supernova-driven bubbles, where compression can promote cooling and star formation.
Extinction maps reveal abundant cold material around the Local Bubble (LB) \citep[e.g.][]{Vergely2022,Edenhofer2024,ONeill2024}, and nearby young star-forming complexes are preferentially located on its surface \citep[e.g.][]{Zucker2022}.
The discovery of cold gas well inside the LB was therefore unexpected and raised the question of how such material can form and survive in a feedback-dominated environment.
The Local Leo Cold Cloud (LLCC), which forms part of the larger Local Ribbon of Cold Clouds (LRCC), is a striking example of cold gas well inside the LB.

In this work, we address two closely related questions.
First, can LLCC-like gas assemble locally inside the LB, or must it be transported into the bubble as pre-existing cold material?
Second, once formed, can cold diffuse gas survive for several Myr in a warm--hot, feedback-driven environment?
We first summarise the observational constraints on the LLCC and LRCC and review previous theoretical explanations.
In Sects.~\ref{sec:sims}--\ref{sec:cr}, we introduce the \textsc{SILCC} simulations and use a representative snapshot to characterise the morphology and pressure environment of the cold diffuse medium.
In Sects.~\ref{sec:tracking}--\ref{sec:sne}, we extend the analysis over $\sim200$~Myr and use Lagrangian tracer particles to reconstruct how cold material assembles, cools, survives, and disperses.
We discuss the implications for the LLCC and summarise our conclusions in Sect.~\ref{sec:discussion}, with the principal caveats presented in Sect.~\ref{sec:caveats}.

\paragraph{Observational constraints on the LLCC and LRCC}

\citet{Verschuur1969} first identified two distinct patches of cold gas in 21-cm emission, denoted `Cloud A' and `Cloud B', near Galactic longitude $l=230^{\circ}$ and latitude $b=+45^{\circ}$.
Subsequent measurements of their low \ion{H}{1} spin temperatures established both regions as exceptionally cold components of the local interstellar medium (ISM) \citep{Crovisier1985}.
\citet{Heiles2003} showed that the clouds form part of a narrow, discontinuous ribbon of cold gas extending over more than $20^{\circ}$ across the constellation Leo.
At the time, however, their distance remained poorly constrained because these measurements relied on \ion{H}{1} emission and absorption towards extragalactic sources rather than absorption towards stars with known distances.

\citet{Meyer2006} obtained the first high-resolution optical spectra of Clouds A and B, for which the name Local Leo Cold Cloud was subsequently adopted.
They detected narrow interstellar \ion{Na}{1} D absorption towards stars projected behind the 21-cm outline of the cloud.
The detection of this absorption towards stars at known distances placed a firm upper distance limit of 45~pc on the LLCC, establishing it as the nearest known cold cloud to the Sun and locating it well inside the LB.
Subsequent observations further constrained the LLCC to lie between HIP~47513 at 11.3~pc, the most distant star towards the cloud without detected interstellar \ion{Na}{1} absorption, and Regulus at 24.3~pc, the nearest star along these sight lines showing absorption \citep{Peek2011}.
The LLCC is therefore the nearest known cold cloud containing neutral hydrogen and metals.

The velocity of the LLCC is similar to the mean velocity of the cluster of local interstellar clouds (CLIC).
Its heliocentric velocity and direction of motion \citep{Haud2010} closely match those of most CLIC clouds \citep{Redfield2008}, as illustrated in Fig.~11 of \citet{Peek2011}.
This kinematic similarity suggests a possible connection between the two structures, although the CLIC extends only to approximately 10~pc and is therefore considerably closer to the Sun than the LLCC.

\citet{Opher2024b} estimate that the Solar System crossed the edge of the LB approximately 7~Myr ago.
If the LLCC already existed at that time, its similar motion may indicate that it entered the LB together with material associated with the CLIC.
\citet{Zucker2025} argue that the present spatial and kinematic structure of the CLIC is consistent with formation in the remnant of a supernova in Upper Centaurus Lupus approximately 1.2~Myr ago, with individual clouds forming progressively over the past $\sim1$~Myr.

The LLCC itself forms part of the substantially larger LRCC, also known as the Local Lynx Cold Cloud.
The LRCC extends over more than $50^{\circ}$ in declination and four hours of right ascension through the constellations Sextans, Leo, Cancer, and Lynx \citep{Haud2010}.
\citet{Opher2024a} find a small but non-negligible probability that the Sun entered the LRCC approximately 2--3~Myr ago.
Such an encounter would have compressed the heliosphere to a heliopause distance of only 0.22~au, potentially exposing the Earth directly to the dense, cold ISM.

Observations consistently indicate that the LLCC contains exceptionally cold atomic gas.
Early measurements yielded \ion{H}{1} spin temperatures of approximately 20 and 14~K for Clouds A and B \citep{Crovisier1985}, while \citet{Heiles2003} measured temperatures of 22 and 17~K, respectively.
From high-resolution \ion{Na}{1} absorption spectra, \citet{Meyer2006} inferred a gas temperature of $T=20^{+6}_{-8}$~K and a turbulent velocity of $v_\mathrm{turb}=0.37\pm0.08$~km~s$^{-1}$.
After accounting for the 1.05~km~s$^{-1}$ hyperfine splitting of the \ion{Na}{1} lines, \citet{Peek2011} derived temperatures of approximately 16--26~K across the cloud.

Using the distance limits from stellar absorption measurements, \citet{Peek2011} inferred a projected length of 2.8--5.9~pc and a width of 0.25--0.54~pc.
Nearly all of the 23 sight lines with detected \ion{Na}{1} absorption from \citet{Meyer2006} lie within the contours of the \ion{H}{1} 21-cm emission mapped by \citet{Heiles2003}, indicating that the optical and radio observations trace the same coherent structure.

Its three-dimensional geometry, however, remains uncertain.
\citet{Heiles2003} initially considered a geometry in which the clouds are thin along the line of sight, whereas \citet{Peek2011} argued that their line-of-sight thickness is more likely comparable to their projected width, favouring a tubular or filamentary structure.
Estimates of the \ion{H}{1} column density also vary substantially between different measurements and sight lines.
\citet{Heiles2003} obtained values of order $N(\mathrm{H\,I})\approx3\times10^{19}$~cm$^{-2}$ from 21-cm observations, with column densities of $1.7\times10^{19}$ and $3.2\times10^{19}$~cm$^{-2}$ towards Clouds A and B, respectively.
\citet{Peek2011} found column densities reaching approximately $2.5\times10^{20}$~cm$^{-2}$.

For comparison, individual CLIC clouds typically have $N(\mathrm{H\,I})\approx10^{18}$~cm$^{-2}$ \citep{Redfield2008}.
The LLCC therefore has a substantially larger neutral-gas column than the nearby CLIC clouds.
Its column density is instead comparable to that associated with the boundary of the LB, which begins near $N(\mathrm{H\,I})\approx2\times10^{19}$~cm$^{-2}$ and reaches characteristic values of approximately $3\times10^{20}$~cm$^{-2}$ \citep{Peek2011}.

Assuming a tube-like geometry with a line-of-sight thickness comparable to the projected width of 0.25--0.54~pc, \citet{Peek2011} derived an \ion{H}{1} number density of 150--320~cm$^{-3}$ and a thermal pressure of $P_\mathrm{th,LLCC}/k_\mathrm{B}=2250$--$9600$~K~cm$^{-3}$.
A thinner, sheet-like geometry would imply a larger density and thermal pressure.

The \ion{C}{1} fine-structure lines provide an independent diagnostic of the density and pressure of the LLCC.
The excited $J=1$ and $J=2$ levels lie at $E/k_\mathrm{B}=23.6$ and 62.4~K above the ground state, respectively.
Their relative populations are set mainly by collisions with atomic hydrogen and depend differently on temperature and density.
The method developed by \citet{Jenkins2001} uses the ratios $f_1=N(\mathrm{C\,I}^{*})/N(\mathrm{C\,I})_\mathrm{tot}$ and $f_2=N(\mathrm{C\,I}^{**})/N(\mathrm{C\,I})_\mathrm{tot}$ to infer the density, and hence the pressure, when an independent temperature constraint is available.
Applying this method to 89 sight lines, \citet{Jenkins2011} found a mean thermal pressure of $P_\mathrm{th}/k_\mathrm{B}=3800$~K~cm$^{-3}$ for the approximately log-normal pressure distribution.
A small fraction of the gas instead reaches $P_\mathrm{th}/k_\mathrm{B}\geq3\times10^5$~K~cm$^{-3}$ at $T>80$~K.
These high pressures correlate with large turbulent velocities and may be produced by supernova shocks.

Using the same \ion{C}{1} population-ratio technique, \citet{Meyer2012} inferred an average thermal pressure of $P_\mathrm{th,LLCC}/k_\mathrm{B}\approx6\times10^4$~K~cm$^{-3}$ from two sight lines.
For an assumed kinetic temperature of 20~K, this pressure corresponds to an \ion{H}{1} number density of approximately $3\times10^3$~cm$^{-3}$.
For this density and a representative column density of $N(\mathrm{H\,I})\approx10^{19}$~cm$^{-2}$, the implied thickness is approximately 200~au, consistent with a thin sheet or filament.
\citet{Meyer2012} did not quote an uncertainty range for the LLCC thermal pressure.
However, \citet{Jenkins2011} showed that mixtures of gas densities or radiation fields along a sight line can bias the inferred pressure.
Additional sight lines through the LLCC may therefore lead to a revised estimate.

If $P_\mathrm{th,LLCC}/k_\mathrm{B}\approx6\times10^4$~K~cm$^{-3}$, the surrounding pressure must be considered when assessing the stability of the cloud.
This value exceeds the estimated total pressure from thermal gas, magnetic fields, cosmic rays (CRs), and ram pressure in the Local Interstellar Cloud, $P_\mathrm{tot,LIC}/k_\mathrm{B}\approx2.36\times10^4$~K~cm$^{-3}$.
It also exceeds the estimated total pressure of a LB filled with warm Str\"omgren-sphere gas, $P_\mathrm{tot,warm\,LB}/k_\mathrm{B}\approx1.96\times10^4$~K~cm$^{-3}$ \citep{Linsky2023}, and the vertical gravitational pressure near the Galactic plane, $P_\mathrm{grav}/k_\mathrm{B}\approx2.2\times10^4$~K~cm$^{-3}$ \citep{Cox2005}.
By contrast, the estimated total pressure of a LB filled with million-degree plasma, $P_\mathrm{tot,hot\,LB}/k_\mathrm{B}\approx4.64\times10^4$~K~cm$^{-3}$ \citep{Linsky2023}, is comparable to the inferred LLCC thermal pressure.

Finally, the \ion{H}{1} 21-cm emission is best fitted by two Gaussian velocity components.
\citet{Peek2011} favoured a configuration in which the more opaque component lies in front of the second component.
The two components approach each other at approximately 0.4~km~s$^{-1}$.

\paragraph{Previous theoretical explanations for cold gas inside the Local Bubble}

The processes responsible for forming and maintaining cold gas inside the LB remain uncertain.
Existing models have considered collisions between warm clouds or converging warm-gas flows, which compress material into a dense interface that subsequently cools radiatively.
Hydrodynamic simulations of converging diffuse warm gas indeed produced shocked interface layers containing cold material \citep[e.g.][]{Audit2005,Heitsch2006,VazquezSemadeni2006}.

For example, \citet{VazquezSemadeni2006} found that weakly transonic flows can produce a cold layer after approximately 1~Myr.
The resulting layer has $N(\mathrm{H\,I})\approx2.5\times10^{19}$~cm$^{-2}$, a thickness of approximately 6000~au, a temperature near 25~K, and a thermal pressure of approximately 6650~K~cm$^{-3}$.
Its temperature and column density are similar to those inferred by \citet{Meyer2012}, whereas its pressure is lower by approximately one order of magnitude and its thickness is larger by substantially more than one order of magnitude.
The models further predicted that the cold gas was thermally overpressured by a factor of 1.5--4 relative to the surrounding warm gas and fragmented into thin sheets with a `filamentary honeycomb pattern' \citep{VazquezSemadeni2006}.
This morphology is qualitatively consistent with the 21-cm emission and with spatial variations inferred from time-variable \ion{Na}{1} absorption towards the high-proper-motion star HD~84937.
In these models, the predicted line widths of approximately 1~km~s$^{-1}$ primarily traced the inflow velocity rather than internal turbulence.
The simulations of \citet{Audit2005} and \citet{Heitsch2006} further demonstrated how turbulence and inflow strength regulate cold-gas formation.
Across all three studies, thermally unstable gas fragmented into filamentary cold structures.

Another proposed formation channel is fragmentation of the interaction shell between the LB and Loop~I through a hydromagnetic Rayleigh--Taylor instability \citep{Breitschwerdt2000}.
These models demonstrated that cold structures can arise from dynamically compressed warm material, but they did not establish which pathway produced the LLCC or whether such structures can survive for several Myr in a fully feedback-regulated multiphase ISM.

\section{The \textsc{SILCC} simulation and its cold diffuse medium}\label{sec:sims}

\begin{figure}
  \centering
  \includegraphics[width=.95\linewidth]{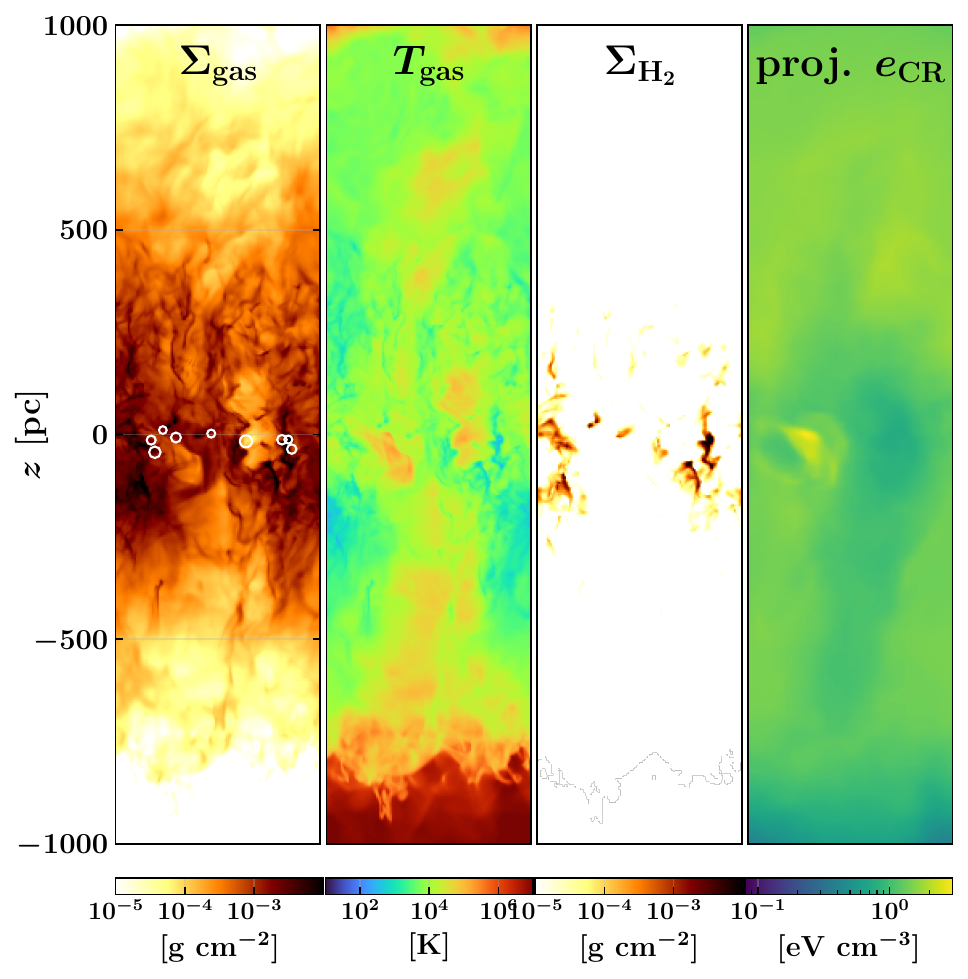}
  \caption{Overview of the \textsc{SILCC} ISM simulation with an initial gas surface density of $\Sigma_\mathrm{init}=10\,M_\odot\,\mathrm{pc}^{-2}$ under solar-neighbourhood conditions.
  The panels show edge-on projections of the total gas column density, $\Sigma_\mathrm{gas}$, the density-weighted gas temperature, $T_\mathrm{gas}$, the molecular hydrogen column density, $\Sigma_\mathrm{H_2}$, and the CR energy density, $e_\mathrm{CR}$.
  White circles in the gas column-density panel mark massive stellar clusters represented by sink particles.
  Further details of the simulation are given by \citet{Rathjen2025}.}
  \label{fig:overview}
\end{figure}

Large-scale multiphysics simulations of the ISM now make it possible to study cold-gas formation and survival in a self-consistent, feedback-regulated environment.
The \textsc{SILCC} Project \citep{Walch2015,Brugaletta2025b} models a representative patch of a vertically stratified galactic disc with magnetohydrodynamic (MHD) turbulence, self-gravity, non-equilibrium chemistry, and self-consistent star formation.
Its time-dependent stellar-feedback model includes stellar winds, radiative transfer of far-ultraviolet (FUV) and extreme-ultraviolet (EUV) radiation, clustered supernovae, and CR acceleration and anisotropic transport.
Cosmic rays are represented as a relativistic fluid in the advection--diffusion approximation and contribute a pressure, $P_\mathrm{CR}$, and energy density, $e_\mathrm{CR}$, to the MHD equations.
The CR fluid has an adiabatic index of $\gamma_\mathrm{CR}=4/3$ and loses energy through adiabatic expansion and hadronic interactions \citep{Pfrommer2017}.
The simulations do not resolve the CR energy spectrum and instead adopt a `grey' treatment with a single effective energy bin represented by 1~GeV protons.
This approximation does not capture the broad, approximately power-law CR spectrum, which spans more than 12 orders of magnitude in particle energy.
At GeV energies, CRs carry much of the dynamically relevant pressure, whereas lower-energy CRs contribute more strongly to processes such as gas heating and ionisation \citep{Girichidis2022,Girichidis2024}.
Our model should therefore be interpreted as an effective description of the dynamically important GeV CR population rather than of the full CR spectrum.

Because diffusive shock acceleration is unresolved, each supernova injects $10^{50}$~erg into the CR fluid, corresponding to 10~per cent of the canonical supernova energy \citep{Helder2012,Ackermann2013}.
The CRs diffuse preferentially along the local magnetic field with fixed coefficients $\kappa_\parallel=10^{28}$~cm$^2$~s$^{-1}$ and $\kappa_\perp=10^{26}$~cm$^2$~s$^{-1}$ \citep[see][and references therein]{Rathjen2021}.
This anisotropy represents pitch-angle scattering by Alfv\'en waves \citep[][and references therein]{Ruszkowski2023}.
The diffusion coefficient, $\kappa$, scales inversely with the total scattering frequency, which itself is indirectly affected by the magnetic-field strength, $|\mathbf{B}|$.
In the simulation, $\kappa_\parallel$ and $\kappa_\perp$ are fixed and do not vary with the local magnetic-field strength.
The magnetic field affects CR transport through the orientation of the anisotropic diffusion tensor, so the tangled field geometry produces a spatially complex CR distribution even with constant diffusion coefficients.

Our analysis focuses on the solar-neighbourhood simulation previously published as $\Sigma010$vFUV in \citet{Rathjen2025}.
It has an initial gas surface density of $\Sigma_\mathrm{init}=10\,M_\odot\,\mathrm{pc}^{-2}$, an initial horizontal magnetic field of $B_x=6~\mu$G, and a maximum spatial resolution of $\Delta x\approx3.9$~pc.
The simulation produces hot supernova remnants, ionised regions near $10^4$~K, dense star-forming gas, and a cold diffuse component far from young stellar sources.
Although it does not model the LB directly, its parsec-scale resolution captures the multiphase structure and feedback processes needed to test whether cold gas can exist in a Local-Bubble-like environment.
Figure~\ref{fig:overview} shows a representative overview of the simulation.

\citet{Rathjen2025} find that a cold diffuse medium (CDM), defined by $T<100$~K and $0<n_\mathrm{H_2}\leq2$~cm$^{-3}$, occurs preferentially far from young star-forming sites, where the gas is exposed mainly to the weak background FUV radiation field.
The resulting reduction in photoelectric heating and photodissociation permits a cold molecular component at low density.
The cold diffuse branch is strongly suppressed in an otherwise identical simulation without CR acceleration and transport, indicating that CR pressure is an important component of its dynamical support.

\begin{figure}
  \centering
  \includegraphics[width=.95\linewidth]{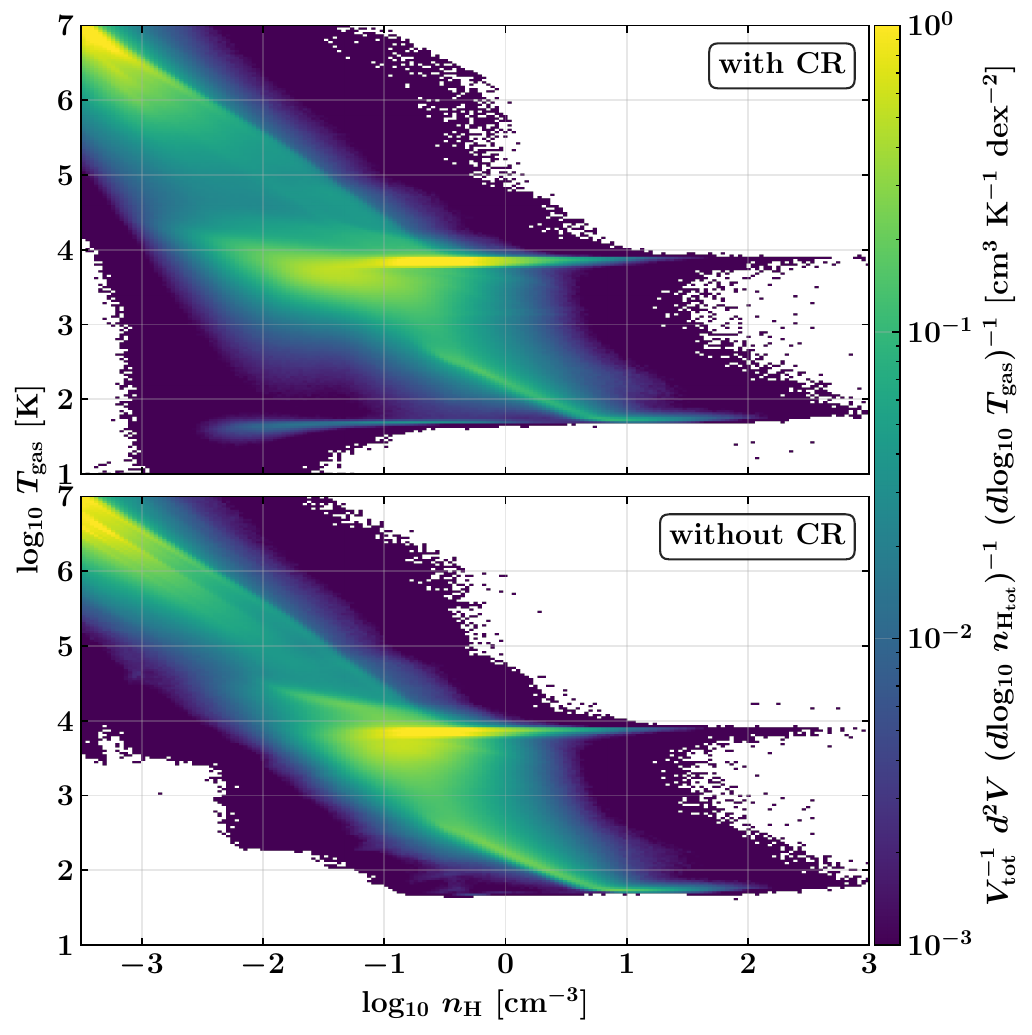}
  \caption{Volume-weighted joint probability density functions showing the gas distribution in the galactic plane under solar-neighbourhood conditions as a function of gas temperature $T_\mathrm{gas}$ and total hydrogen number density $n_\mathrm{H}$.
  The simulation in the upper panel includes CR acceleration and transport, whereas the simulation in the lower panel omits CRs.
  Otherwise, the initial conditions and included physical processes are identical between the two models.
  The cold diffuse branch is strongly suppressed when CR pressure is omitted.}
  \label{fig:phase}
\end{figure}

Figure~\ref{fig:phase} shows the volume-weighted joint probability density function, $\mathrm{PDF}_\mathrm{vw}$, of gas temperature, $T_\mathrm{gas}$, and hydrogen-nuclei number density, $n_\mathrm{H}$, for otherwise identical solar-neighbourhood models with CR transport (upper panel) and without CRs (lower panel).
We combined the full evolution after the onset of star formation in 1~Myr bins, extending to 200 and 130~Myr for the two models, respectively.
Both simulations contain hot gas produced by overpressured supernova remnants, a characteristic \ion{H}{2}-region branch near $T\approx10^4$~K, and cold, dense gas reaching $n_\mathrm{H}\approx10^3$~cm$^{-3}$.
At the maximum spatial resolution of $\Delta x\approx3.9$~pc, substantially higher densities are not resolved.
When the gas exceeds this threshold, and further star-formation checks are met, it is converted into star cluster sink particles as part of our star formation subgrid model \citep[see][]{Gatto2017}.

The most conspicuous difference introduced by CRs is the appearance of a cold, diffuse gas phase.
In the selected snapshot, the CDM has a mean separation of $d_\mathrm{OB}=221\pm131$~pc from the nearest OB star cluster.
The large distance from young stellar sources reduces the local interstellar radiation field (ISRF), photoelectric heating, and photodissociation.
The reduced radiative heating allows the CDM to cool to temperatures of only a few tens of kelvin.
We next examine how this cold phase is structured and how it is supported against its much warmer surroundings.

\subsection{Morphology of the CDM}\label{sec:morphology}

\begin{figure}
  \centering
  \includegraphics[width=.95\linewidth]{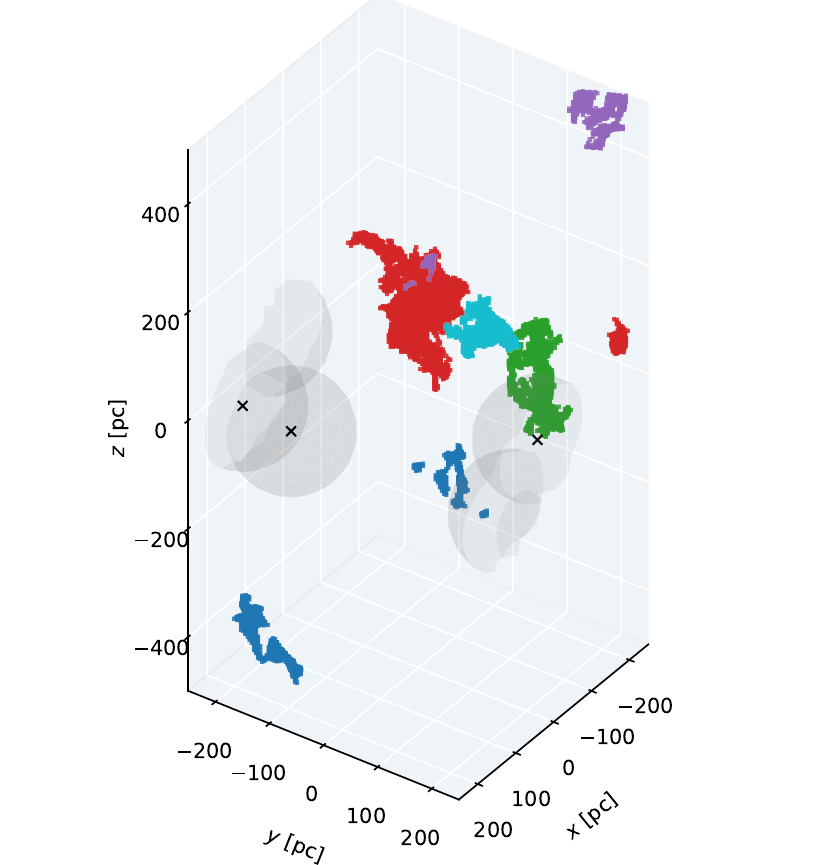}
  \includegraphics[width=.95\linewidth]{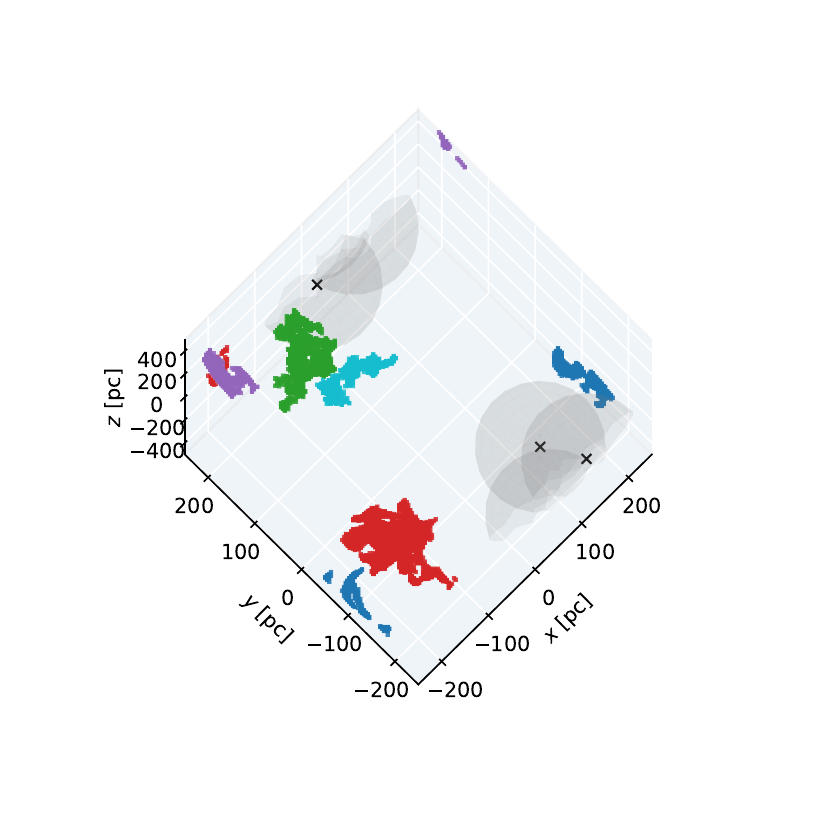}
  \caption{Five largest connected regions satisfying $T<100$~K and $0<n_\mathrm{H_2}\leq2$~cm$^{-3}$ at $t\approx100$~Myr.
  The colours identify the five friends-of-friends groups listed in Table~\ref{tab:overview}.
  Black crosses mark active sink particles containing at least one living massive star.
  The translucent grey surfaces show spheres with radii of 100~pc around these sinks for visual reference.
  The CDM is predominantly filamentary or sheet-like rather than compact.
  Periodic boundary conditions are applied along the $x$- and $y$-directions.}
  \label{fig:morphology}
\end{figure}

\begin{table*}
  \centering
  \caption{Properties of the five largest connected regions satisfying $T<100$~K and $0<n_\mathrm{H_2}\leq2$~cm$^{-3}$ at $t\approx100$~Myr.}
  \setlength{\tabcolsep}{2pt}
  \begin{tabular}{lcccccccc|cccr}
  \hline
  Group
   & $T_\mathrm{in}$ 
   & $T_\mathrm{out}$ 
   & $|\mathbf{B}|_\mathrm{in}$ 
   & $|\mathbf{B}|_\mathrm{out}$ 
   & $P_\mathrm{th,in}/k_\mathrm{B}$ 
   & $P_\mathrm{th,out}/k_\mathrm{B}$ 
   & $P_\mathrm{CR,in}/k_\mathrm{B}$ 
   & $P_\mathrm{CR,out}/k_\mathrm{B}$ 
   & $\lambda_1$ 
   & $\lambda_2$ 
   & $\lambda_3$ 
   & Classification \\
   & 
  [K]
   & [$10^3$ K] 
   & [$\mu$G] 
   & [$\mu$G] 
   & [$10^2$ K~cm$^{-3}$] 
   & [$10^2$ K~cm$^{-3}$] 
   & [$10^3$ K~cm$^{-3}$] 
   & [$10^3$ K~cm$^{-3}$] 
   & [pc$^2$] 
   & [pc$^2$] 
   & [pc$^2$] 
   & \\
  \hline
  \hline
  red
   & $52^{54}_{50}$ 
   & $0.93^{4.46}_{0.11}$ 
   & $4.0^{6.3}_{2.0}$ 
   & $2.3^{4.4}_{0.8}$ 
   & $5.54^{11.45}_{1.73}$ 
   & $10.3^{25.4}_{2.9}$ 
   & $12.56^{14.67}_{11.29}$ 
   & $12.04^{14.96}_{10.96}$ 
   & 3612 
   & 1358 
   & 514 
   & mixed/triaxial \\
  green
   & $53^{54}_{52}$ 
   & $0.57^{8.32}_{0.09}$ 
   & $3.8^{5.6}_{2.2}$ 
   & $2.7^{4.6}_{0.9}$ 
   & $4.49^{7.91}_{0.36}$ 
   & $5.2^{24.4}_{1.9}$ 
   & $11.07^{12.53}_{10.49}$ 
   & $11.33^{13.07}_{10.41}$ 
   & 2515 
   & 301 
   & 91 
   & filament \\
  blue
   & $46^{51}_{12}$ 
   & $2.56^{8.73}_{0.79}$ 
   & $0.2^{0.3}_{0.1}$ 
   & $0.2^{0.6}_{0.1}$ 
   & $0.02^{0.03}_{0.00}$ 
   & $1.3^{4.7}_{0.4}$ 
   & $10.12^{10.33}_{9.92}$ 
   & $10.18^{10.56}_{10.00}$ 
   & 1019 
   & 733 
   & 57 
   & sheet \\
  cyan
   & $53^{54}_{52}$ 
   & $0.99^{934.51}_{0.11}$ 
   & $3.5^{4.8}_{2.0}$ 
   & $1.9^{3.5}_{0.2}$ 
   & $4.77^{9.49}_{0.34}$ 
   & $4.8^{14.5}_{2.1}$ 
   & $12.63^{12.93}_{12.21}$ 
   & $12.50^{12.98}_{11.64}$ 
   & 1011 
   & 208 
   & 106 
   & filament \\
  purple
   & $18^{44}_{10}$ 
   & $6.63^{13.04}_{2.37}$ 
   & $0.4^{0.6}_{0.2}$ 
   & $0.4^{0.7}_{0.1}$ 
   & $0.002^{0.007}_{0.001}$ 
   & $0.9^{2.2}_{0.3}$ 
   & $11.15^{11.65}_{10.80}$ 
   & $11.16^{11.79}_{10.79}$ 
   & 1019 
   & 503 
   & 27 
   & sheet \\
  \hline
  \end{tabular}
  \tablefoot{The colours refer to the friends-of-friends groups shown in Fig.~\ref{fig:morphology}.
  We list the temperature, magnetic-field strength, thermal pressure, and CR pressure inside each CDM group and in its surrounding ambient shell.
  The shell comprises the three immediately adjacent grid-cell layers around each group.
  Each entry lists the median, with the 84th and 16th percentiles given as the superscript and subscript, respectively.
  The final columns give the eigenvalues of the principal-component analysis and the resulting morphological classification.}  
  \label{tab:overview}
\end{table*}

We used a representative snapshot to examine the morphology and immediate environment of the CDM in detail.
The snapshot represents solar-neighbourhood conditions with $\Sigma_\mathrm{init}=10\,M_\odot\,\mathrm{pc}^{-2}$ and an initially uniform horizontal magnetic field of $B_x=6~\mu$G.
It is taken at $t\approx100$~Myr.
During the preceding 10~Myr, the mean star formation rate surface density is $\Sigma_\mathrm{SFR}=(1.2\pm0.4)\times10^{-3}\,M_\odot\,\mathrm{yr}^{-1}\,\mathrm{kpc}^{-2}$.
During this interval, 92 core-collapse supernovae occurred within the central $(500\times500\times1000)$~pc$^3$ volume.
At the selected time, the midplane ISM contains 14 active O-type stars with $M_\star\geq16\,M_\odot$ and 103 active B-type stars with $M_\star>8\,M_\odot$.
These stars provide stellar-wind, FUV, and EUV feedback and will eventually produce supernovae.

We identified connected CDM regions with a friends-of-friends (FoF) algorithm following the general approach of \citet{Davis1985}.
We mapped the simulation onto a uniform grid with $\Delta x\approx3.9$~pc and selected cells satisfying $T<100$~K and $0<n_\mathrm{H_2}\leq2$~cm$^{-3}$ as CDM.
The upper density threshold excludes dense molecular clouds from the selected population.
We linked the selected cells through shared faces and edges (i.e. 18-connectivity), corresponding to a linking length of $r_\mathrm{l}=\sqrt{2}\,\Delta x$.
We required at least 125 cells per FoF group to reduce the sensitivity of the catalogue to numerical noise.

Figure~\ref{fig:morphology} shows the five FoF groups with the largest volumes.
For each group, we evaluated the median temperature, $T$, magnetic-field strength, $|\mathbf{B}|$, thermal pressure, $P_\mathrm{th}$, and CR pressure, $P_\mathrm{CR}$.
We report the median, with the 84th and 16th percentiles given as the superscript and subscript, respectively.
The subscripts `in' and `out' in Table~\ref{tab:overview} denote the FoF cells and the surrounding ambient shell, respectively.
We defined the ambient shell to extend to $r_\mathrm{out}=3\Delta x$ beyond the FoF boundary.

The red, green, and cyan groups have comparatively narrow internal temperature distributions.
The blue group has a broader and colder distribution, with $T_\mathrm{in}=46^{51}_{12}$~K.
The purple group contains the coldest gas among the five selected structures, with $T_\mathrm{in}=18^{44}_{10}$~K.

The ambient shells are substantially warmer than the interiors, although none of the groups has a median external temperature of order $10^5$~K.
The external median temperatures range from $T_\mathrm{out}=0.57^{8.32}_{0.09}\times10^3$~K for the green group to $T_\mathrm{out}=6.63^{13.04}_{2.37}\times10^3$~K for the purple group.
The red shell has $T_\mathrm{out}=0.93^{4.46}_{0.11}\times10^3$~K.
The cyan shell is strongly multiphase, with $T_\mathrm{out}=0.99^{934.51}_{0.11}\times10^3$~K.
Its median remains near $10^3$~K, whereas its 84th percentile reaches almost $10^6$~K, showing that a substantial part of the sampled shell extends into the hot phase, consistent with gas heated by stellar winds and supernova feedback.
Except for the red group, the external 84th percentile temperature exceeds $8\times 10^3$~K, typical of \ion{H}{2} regions, indicating that the CDM is embedded in active stellar feedback regions.

The thermal-pressure contrast varies strongly between the five structures.
For the red group, the median thermal pressure changes from $P_\mathrm{th,in}/k_\mathrm{B}=5.54^{11.45}_{1.73}\times10^2$~K~cm$^{-3}$ inside to $P_\mathrm{th,out}/k_\mathrm{B}=10.3^{25.4}_{2.9}\times10^2$~K~cm$^{-3}$ outside.
The green and cyan groups likewise have internal and external median thermal pressures of the same order of magnitude.
In contrast, the blue group has $P_\mathrm{th,in}/k_\mathrm{B}=0.02^{0.03}_{0.00}\times10^2$~K~cm$^{-3}$ and $P_\mathrm{th,out}/k_\mathrm{B}=1.3^{4.7}_{0.4}\times10^2$~K~cm$^{-3}$.
The purple group shows an even stronger contrast, with $P_\mathrm{th,in}/k_\mathrm{B}=0.002^{0.007}_{0.001}\times10^2$~K~cm$^{-3}$ and $P_\mathrm{th,out}/k_\mathrm{B}=0.9^{2.2}_{0.3}\times10^2$~K~cm$^{-3}$.

The magnetic-field strength generally remains of the same order of magnitude inside and outside a given structure.
For the red group, the field changes from $|\mathbf{B}|_\mathrm{in}=4.0^{6.3}_{2.0}~\mu$G to $|\mathbf{B}|_\mathrm{out}=2.3^{4.4}_{0.8}~\mu$G.
The purple group has substantially weaker fields, with $|\mathbf{B}|_\mathrm{in}=0.4^{0.6}_{0.2}~\mu$G and $|\mathbf{B}|_\mathrm{out}=0.4^{0.7}_{0.1}~\mu$G.

The CR pressure distributions show substantially less variation across the FoF boundaries.
For example, the red group has $P_\mathrm{CR,in}/k_\mathrm{B}=12.56^{14.67}_{11.29}\times10^3$~K~cm$^{-3}$ and $P_\mathrm{CR,out}/k_\mathrm{B}=12.04^{14.96}_{10.96}\times10^3$~K~cm$^{-3}$.
The purple group similarly has $P_\mathrm{CR,in}/k_\mathrm{B}=11.15^{11.65}_{10.80}\times10^3$~K~cm$^{-3}$ and $P_\mathrm{CR,out}/k_\mathrm{B}=11.16^{11.79}_{10.79}\times10^3$~K~cm$^{-3}$.
The CR component therefore dominates over thermal pressure in all five structures, particularly in the blue and purple groups, and substantially reduces the inside--outside contrast once thermal, magnetic, and CR pressure are summed.

A principal-component analysis of the cell coordinates yields eigenvalues $\lambda_1\geq\lambda_2\geq\lambda_3$, corresponding to the coordinate variances along the three principal axes.
Approximately equal eigenvalues indicate an approximately isotropic structure; one dominant eigenvalue is characteristic of a filament; and two dominant eigenvalues are characteristic of a sheet (compact: $\lambda_1\approx\lambda_2\approx\lambda_3$; filamentary: $\lambda_1\gg\lambda_2\approx\lambda_3$; sheet-like: $\lambda_1\approx\lambda_2\gg\lambda_3$).
All five structures have $\lambda_3\ll\lambda_1$, indicating strongly anisotropic geometries.
We did not impose strict axis-ratio thresholds, but instead assigned descriptive morphological classes from the relative eigenvalues listed in Table~\ref{tab:overview}.
All five structures are inconsistent with approximately isotropic compact clouds and instead show filamentary, sheet-like, or strongly triaxial morphologies that qualitatively resemble the extended geometry of the LLCC and LRCC.
However, the $3.9$~pc cell size does not resolve the observed 0.25--0.54~pc width of the LLCC.
The simulated FoF structures should therefore be interpreted as resolved patches or complexes of the cold diffuse phase rather than as one-to-one analogues of the observed cloud.

\subsection{The role of cosmic rays}\label{sec:cr}

As described in Sect.~\ref{sec:sims}, CRs provide a non-thermal pressure component that is transported anisotropically throughout the simulated ISM.
Their comparatively long loss times in diffuse gas allow CR energy to spread over large regions and influence the dynamical and pressure structure of the multiphase ISM \citep[e.g.][]{Girichidis2018,Rathjen2021,Rathjen2023,Shimoda2025}.
They can also regulate star and star-cluster formation \citep[see e.g.][]{Sike2026}.
On larger scales, the resulting CR pressure gradient can support and accelerate multiphase galactic outflows \citep[see e.g.][]{Rathjen2021,Chan2022,Girichidis2022,Rathjen2023,Simpson2023}.
Here, we focus on how this distributed CR pressure modifies the balance between the CDM structures and their surroundings.

\begin{figure}
  \centering
  \includegraphics[width=.95\linewidth]{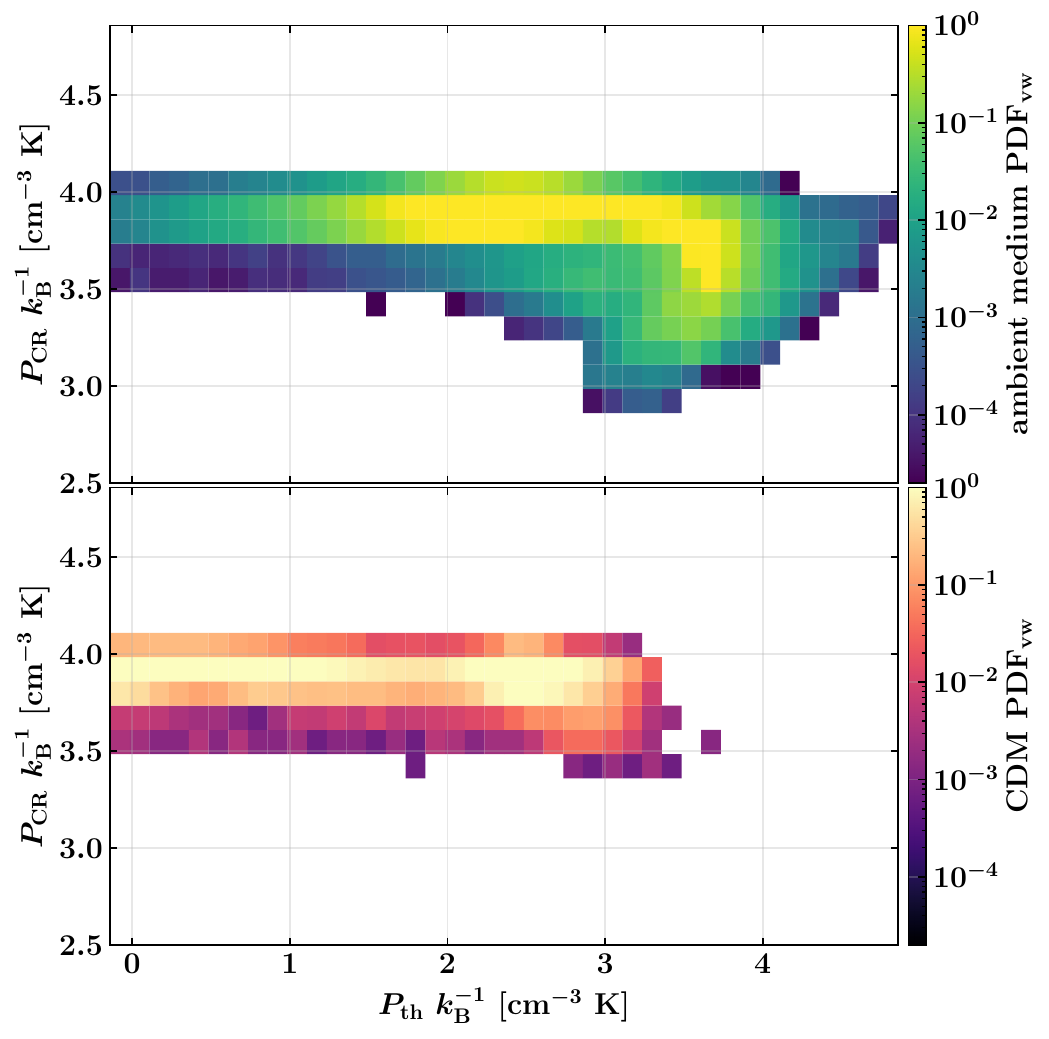}
  \caption{Volume-weighted joint probability density functions, $\mathrm{PDF}_\mathrm{vw}$, of the CR pressure, $P_\mathrm{CR}k_\mathrm{B}^{-1}$, and thermal pressure, $P_\mathrm{th}k_\mathrm{B}^{-1}$, of the CDM (bottom) and the ambient medium (top).
  While the CDM resides at low thermal pressure, CR pressure provides a shared non-thermal pressure floor across the cold-gas boundary and reduces the relative importance of the external thermal-pressure excess.}
  \label{fig:presDist}
\end{figure}

In Fig.~\ref{fig:presDist}, we compare the distributions of thermal pressure, $P_\mathrm{th}$, and CR pressure, $P_\mathrm{CR}$, in the CDM (bottom) and in its surrounding medium (top).
The CDM occupies substantially lower thermal pressures than its surroundings, whereas the CR pressure distributions overlap strongly.
The CR pressure therefore acts primarily as a shared non-thermal pressure floor rather than as a systematic pressure enhancement confined to the cold gas.
Together with magnetic pressure, it reduces the inside--outside pressure contrast relative to that inferred from thermal pressure alone.
The comparatively narrow range of $P_\mathrm{CR}$ sampled by the CDM is consistent with a spatially smoother CR energy distribution than the strongly structured thermal gas.
The corresponding median pressure components of the five example structures are listed in Table~\ref{tab:overview}.

\section{Origin, cooling, and survival of cold material}\label{sec:tracking}

We extended the analysis from a representative snapshot to the full time evolution to determine how CDM structures assemble, cool, and survive.
For this purpose, we applied the FoF analysis to the solar-neighbourhood simulation over $t\approx30$--$200$~Myr\footnote{We excluded the first 30~Myr because this interval is dominated by the initial-condition transient and precedes the established feedback-regulated state.}.
We additionally used massless Lagrangian tracer particles to reconstruct the thermal and spatial histories of the material associated with individual CDM structures.
The tracers are advected with the gas on the Eulerian grid and record the properties of the cells in which they reside.
For this time-dependent analysis, we retained the molecular-density criterion $0<n_\mathrm{H_2}\leq2$~cm$^{-3}$, the 18-neighbour connectivity, and the 125-cell minimum size used above, but adopted the stricter threshold $T<55$~K to select gas closer in temperature to the observed LLCC.

We identified connected structures in the central $(500\times500\times1000)$~pc$^3$ volume on a $128\times128\times256$ grid with $\Delta x=3.9$~pc and linked their tracer overlap between consecutive outputs.
We required a persistent FoF identity to be detected in at least three consecutive snapshots, corresponding to $\Delta t=0.3$~Myr.
Across all analysed outputs, the catalogue contains 47{,}045 snapshot-level FoF nodes.
Linking these nodes through their tracer overlap yields 8{,}974 tracer-linked tracks, of which 3{,}138 are detected in at least three consecutive snapshots and are therefore classified as persistent.

As a robustness test, we also constructed a purely Eulerian temporal linkage that uses direct overlap of FoF cells together with a periodic centre-of-mass proximity criterion and contains no tracer information.
This alternative definition yields 5{,}574 linear identities, of which 3{,}341 are persistent.
As expected for structures that fragment, merge, and exchange material, the absolute number of reconstructed identities depends on the temporal-linking definition.
The lifetime distributions, however, are very similar.
The longest-lived identities reach 24.3~Myr with tracer linkage and 23.2~Myr with cell-based linkage.
We retained the tracer-overlap definition as the fiducial tracker because it follows material continuity.
The cell-based comparison shows that the characteristic FoF lifetime distribution is robust to an independent Eulerian continuity criterion and that the transient population with a long-lived tail is not an artefact of the tracer criterion.

Because reconstructing complete tracer histories for every track is computationally prohibitive, we analysed the union of the 150 persistent tracks with the highest peak masses and the 150 with the lowest median temperatures, yielding a selected sample of 300 tracks.
For each track, we sampled up to 250 randomly selected tracer particles at the track's peak mass and followed them at a cadence of 0.5~Myr.
These histories reconstruct the thermal and spatial evolution of the material before it enters and after it leaves the associated CDM structure.
In the following, we refer to this sampled material as `peak-cohort material'.

\subsection{Properties of the CDM}\label{sec:cdm_properties}

We first characterise the full population of identified CDM structures before considering their temporal evolution and survival.
Unless stated otherwise, quoted values give the median and 16th--84th percentile range across all 47{,}045 FoF nodes, using the median gas properties of each structure.

The structures span a broad mass range, with a median mass of $M=84.7^{6.5\times10^{3}}_{8.8}\,M_\odot$.
Their sizes vary less strongly, with a median of 219 cells, corresponding to a spherical-equivalent radius of $R_\mathrm{eq}=14.6^{19.9}_{12.7}$~pc.
The comparatively narrow size distribution is partly imposed by the minimum FoF size of 125 cells, whereas the broad mass distribution reflects the wide range of gas densities within the structures.

The median internal temperature is $T_\mathrm{in}=50.9^{53.3}_{44.0}$~K, whereas the surrounding gas is substantially warmer, with $T_\mathrm{out}=1.16^{4.30}_{0.29}\times10^{3}$~K.
Of all FoF nodes, 18.87~per cent are partially embedded in warm gas ($0.1\leq f_\mathrm{warm,out}<0.5$), while 81.11~per cent are strongly embedded ($f_\mathrm{warm,out}\geq0.5$).
Only nine FoF nodes are not at least partially surrounded by warm gas.
Substantial contact with hot gas, defined by $f_\mathrm{hot,out}\geq0.1$, is less common and occurs for 5.01~per cent of the population.
Warm gas therefore characterises the immediate environment of the CDM, whereas direct contact with a substantial amount of hot gas occurs only in a smaller, but statistically useful, subset.
The structures are typically far from active OB clusters, with a median nearest-cluster distance of $d_\mathrm{OB}=159^{326}_{73}$~pc, but much closer to ionised-gas interfaces, with a median distance of $d_{\mathrm{H\,II}}=14^{55}_{6}$~pc from the nearest \ion{H}{2}-region boundary.

\begin{figure*}
  \centering
  \includegraphics[width=.98\linewidth]{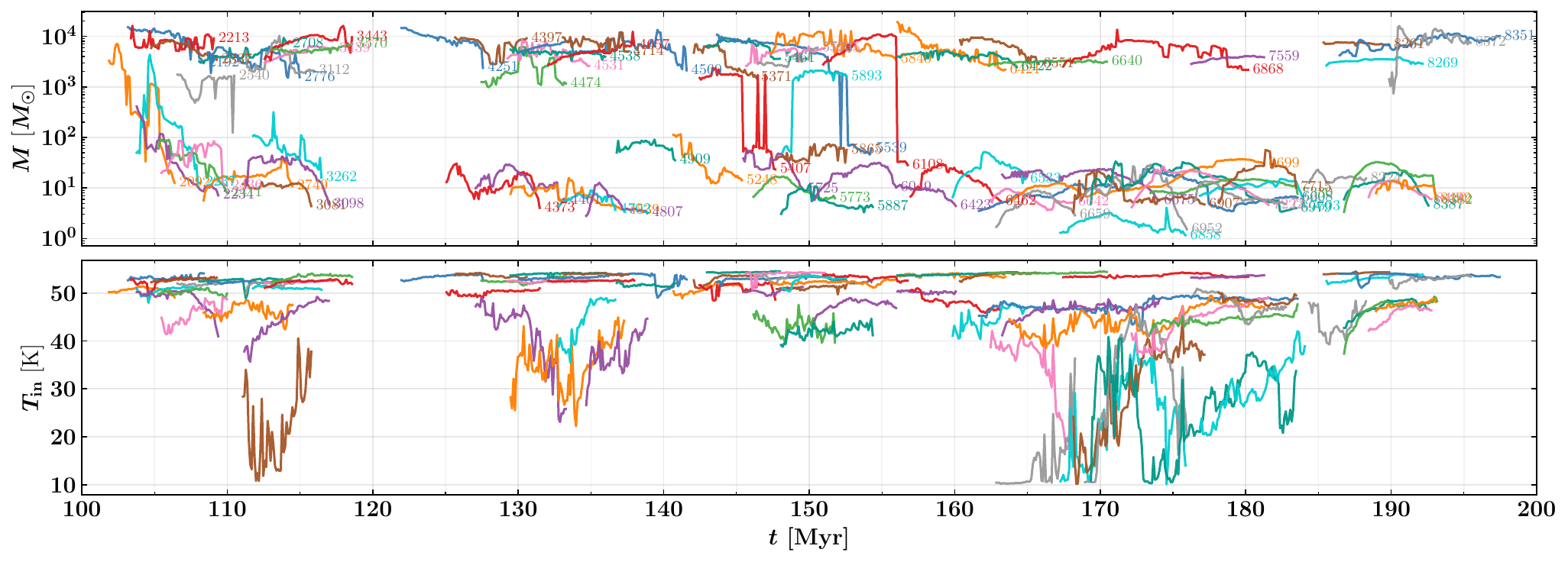}
  \caption{
  Time evolution of CDM structure mass, $M$ (top), and median internal temperature, $T_\mathrm{in}$ (bottom), for selected persistent CDM tracks between $t=100$ and $200$~Myr.
  For clarity, we show tracks containing at least 40 detected nodes ($\Delta t\geq4$~Myr) and no more than ten tracks at a given time.
  Each continuous line represents one tracer-linked FoF identity, labelled by its track identifier.
  }
  \label{fig:cdm_track_population}
\end{figure*}

Figure~\ref{fig:cdm_track_population} illustrates the temporal evolution of representative persistent CDM structures.
Individual identities span masses from a few solar masses to several times $10^{4}\,M_\odot$ and frequently undergo rapid mass changes, occasionally exceeding one order of magnitude over short intervals.
Most tracks remain near $T_\mathrm{in}\approx50$~K, although some temporarily reach median temperatures of only $10$--$40$~K.
The temperature evolution is generally smoother than the mass evolution, and no clear one-to-one relation exists between structure mass and median temperature.

\begin{figure}
  \centering
  \includegraphics[width=.98\linewidth]{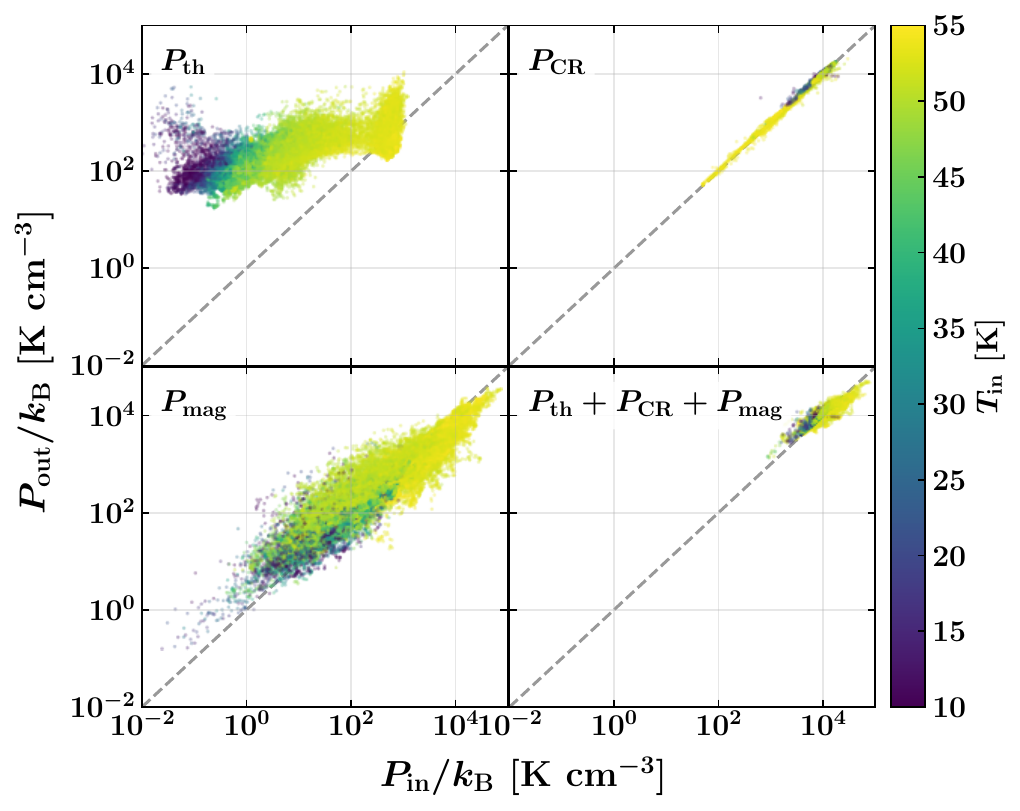}
  \caption{
  Comparison of the median pressure inside each retained CDM FoF node with that in its surrounding three-cell-thick ambient shell.
  The panels show the thermal, CR, magnetic, and summed thermal, CR, and magnetic pressure components.
  Each point represents one FoF node at one simulation output and is coloured by the median internal gas temperature, $T_\mathrm{in}$.
  }
  \label{fig:paired_inside_outside_pressures}
\end{figure}

The temperature contrast between the CDM and its surroundings is accompanied by a strong thermal-pressure contrast.
The median thermal pressure increases from $P_\mathrm{th,in}/k_\mathrm{B}=7.7$~K~cm$^{-3}$ inside the structures to $P_\mathrm{th,out}/k_\mathrm{B}=262$~K~cm$^{-3}$ outside.
By contrast, the magnetic-field strengths are similar, with median values of $|\mathbf{B}|_\mathrm{in}=1.44$~$\mu$G and $|\mathbf{B}|_\mathrm{out}=1.39$~$\mu$G, while the CR pressure is almost continuous across the boundaries, with $P_\mathrm{CR,in}/k_\mathrm{B}=9.12\times10^{3}$ and $P_\mathrm{CR,out}/k_\mathrm{B}=9.14\times10^{3}$~K~cm$^{-3}$.
Consequently, the summed thermal, magnetic, and CR pressures are also similar, with $P_\mathrm{sum,in}/k_\mathrm{B}=1.04\times10^{4}$ and $P_\mathrm{sum,out}/k_\mathrm{B}=1.06\times10^{4}$~K~cm$^{-3}$.

Figure~\ref{fig:paired_inside_outside_pressures} shows that this behaviour extends across the full population.
Thermal pressures lie predominantly above the one-to-one relation, confirming that the CDM is generally thermally underpressured relative to its surroundings.
The CR pressures instead follow the one-to-one relation closely over several orders of magnitude, while magnetic pressure shows the same general behaviour with greater scatter.
The non-thermal components therefore substantially reduce the inside--outside pressure contrast, embedding the thermally distinct CDM in an approximately continuous total-pressure reservoir.

\begin{figure}
  \centering
  \includegraphics[width=.98\linewidth]{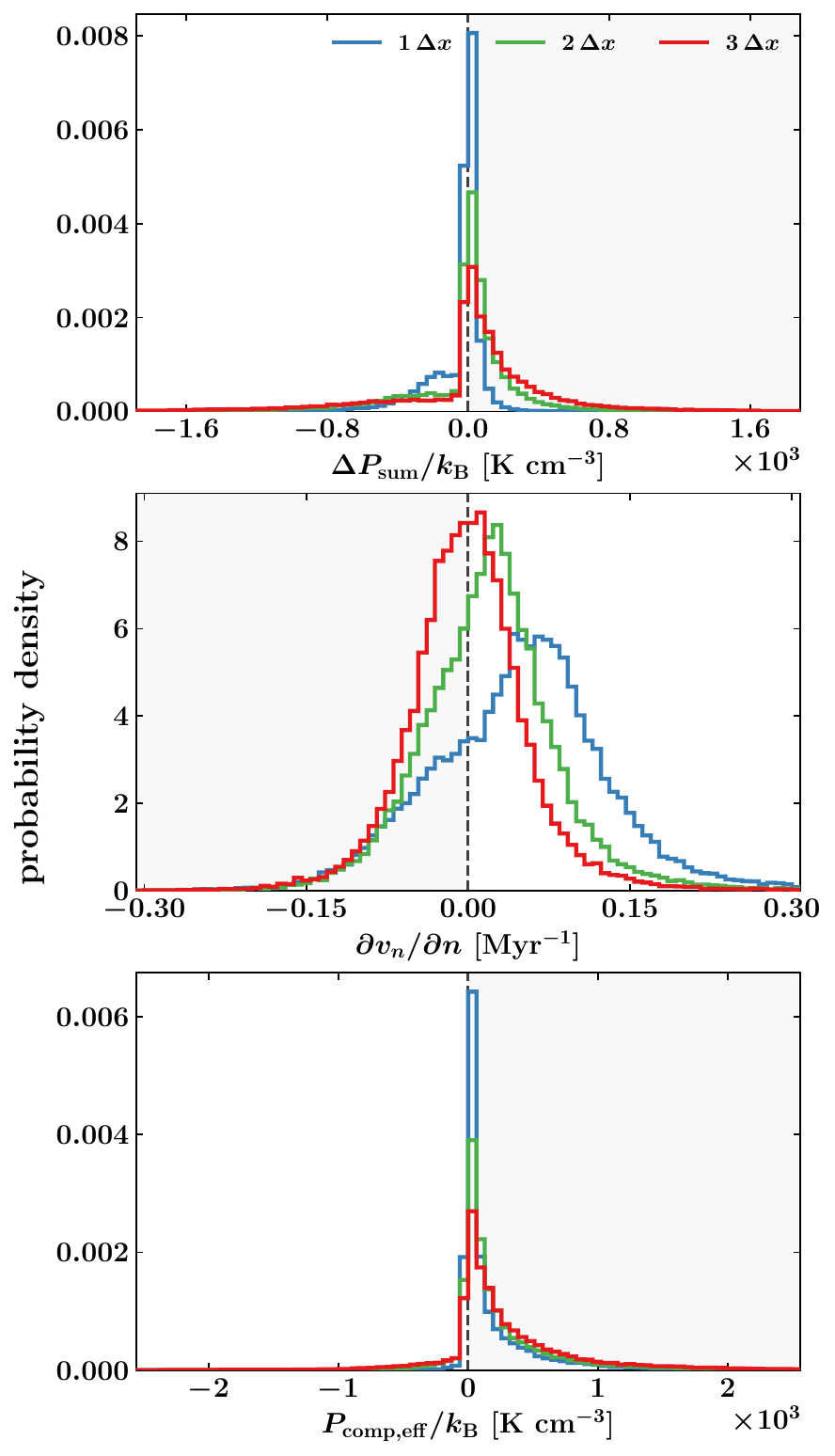}
  \caption{
  Population distributions of the boundary-normal compression diagnostics for all CDM FoF structures, evaluated at separations of one, two, and three grid cells, corresponding to approximately 3.9, 7.8, and 11.7~pc.
  The upper panel shows the median external-minus-internal summed-pressure contrast, $\Delta P_\mathrm{sum}=P_\mathrm{sum,out}-P_\mathrm{sum,in}$, where $P_\mathrm{sum}=P_\mathrm{th}+P_\mathrm{CR}+P_\mathrm{mag}$ and positive values indicate an external pressure excess.
  The middle panel shows the median boundary-normal velocity gradient, $\partial v_n/\partial n=(v_{n,\mathrm{out}}-v_{n,\mathrm{in}})/d$, for which negative values indicate converging flow.
  The lower panel shows the effective compression pressure, $P_\mathrm{comp,eff}=\Delta P_\mathrm{sum}+P_\mathrm{ram,in}$, where $P_\mathrm{ram,in}$ includes only the inward-directed ram-pressure contribution.
  Dashed vertical lines mark zero.
  We first evaluated each quantity across the sampled boundary of an individual FoF structure; the plotted distributions contain the resulting per-structure median values.
  }
  \label{fig:compression_population}
\end{figure}

To determine whether the residual pressure differences are associated with compression or expansion, we evaluated the pressure and velocity changes along the outward normal of each FoF boundary at separations of one, two, and three grid cells.
Figure~\ref{fig:compression_population} shows that the summed-pressure contrast remains strongly concentrated around zero on all three scales.
Its median changes from $\Delta P_\mathrm{sum}/k_\mathrm{B}=-0.4$~K~cm$^{-3}$ at one grid cell to $+20$ and $+50$~K~cm$^{-3}$ at two and three cells, respectively.
At three cells, the 16th--84th percentile range is approximately $-342$ to $+342$~K~cm$^{-3}$, demonstrating substantial structure-to-structure scatter around the small positive median.
These differences are small compared with the characteristic summed pressure of $\sim10^4$~K~cm$^{-3}$, indicating approximate pressure balance with a modest statistical bias towards higher external pressure.

At three-cell separation, the median component contrasts are $\Delta P_\mathrm{th}/k_\mathrm{B}=+95$~K~cm$^{-3}$, $\Delta P_\mathrm{CR}/k_\mathrm{B}=+3$~K~cm$^{-3}$, and $\Delta P_\mathrm{mag}/k_\mathrm{B}=-38$~K~cm$^{-3}$.
The weak positive bias in the summed pressure is therefore driven primarily by thermal pressure, while enhanced internal magnetic pressure partly compensates for it.
The CR pressure contributes little to the local gradient and instead provides the dominant pressure floor on both sides of the interface, upon which the smaller thermal and magnetic differences are superimposed.

Including inward-directed ram pressure increases the median effective compression pressure from approximately $50$ to $105$ and $161$~K~cm$^{-3}$ across the three sampled separations.
The corresponding median fractions of boundary samples with $P_\mathrm{comp,eff}>0$ are 0.62, 0.63, and 0.65.
However, the velocity field does not show a corresponding population-wide converging flow.
The median boundary-normal velocity gradients are $+0.053$, $+0.020$, and $-0.0009$~Myr$^{-1}$, while the median fractions of converging boundary samples increase from 0.37 to 0.45 and 0.50.
Thus, on the largest resolved scale, the population is approximately balanced between converging and diverging boundary motions despite the mild pressure bias towards compression.

The first detected nodes of persistent CDM identities show a somewhat stronger compression bias.
At three-cell separation, their median summed-pressure contrast and effective compression pressure are $+118$ and $+226$~K~cm$^{-3}$, respectively, and a median fraction of 0.69 of their boundary samples satisfies $P_\mathrm{comp,eff}>0$.
Their median velocity gradient nevertheless remains slightly positive at $+0.011$~Myr$^{-1}$, with only 0.48 of the boundary samples converging.
The first appearance of persistent CDM therefore occurs in pressure configurations somewhat more favourable to compression, but without evidence for enhanced resolved convergence at the instant when the FoF criterion is first satisfied.

Structures with substantial hot-gas contact show a clearer dynamical signature.
For these 2{,}349 nodes, the three-cell medians are $\Delta P_\mathrm{sum}/k_\mathrm{B}=+175$~K~cm$^{-3}$, $P_\mathrm{comp,eff}/k_\mathrm{B}=469$~K~cm$^{-3}$, and $\partial v_n/\partial n=-0.030$~Myr$^{-1}$, with a median converging fraction of 0.55.

Overall, these broad distributions show that compression is a statistical tendency rather than a universal dynamical state of the CDM.
Individual structures experience both compression and expansion, with no evidence for a coherent population-wide inflow or a unique collision between warm clouds.
The CDM is instead embedded in a turbulent, approximately pressure-balanced medium in which thermal evolution, local pressure differences, and gas motions continually reshape the cold phase.
In the Appendix, Table~\ref{tab:cdm_population_properties} summarises the principal structural, environmental, and pressure properties of the CDM population, together with the resolved interface diagnostics discussed above.

\subsection{Sensitivity to the magnetic-field strength}\label{sec:mag_sensitivity}

\begin{figure*}
  \centering
  \includegraphics[width=.95\textwidth]{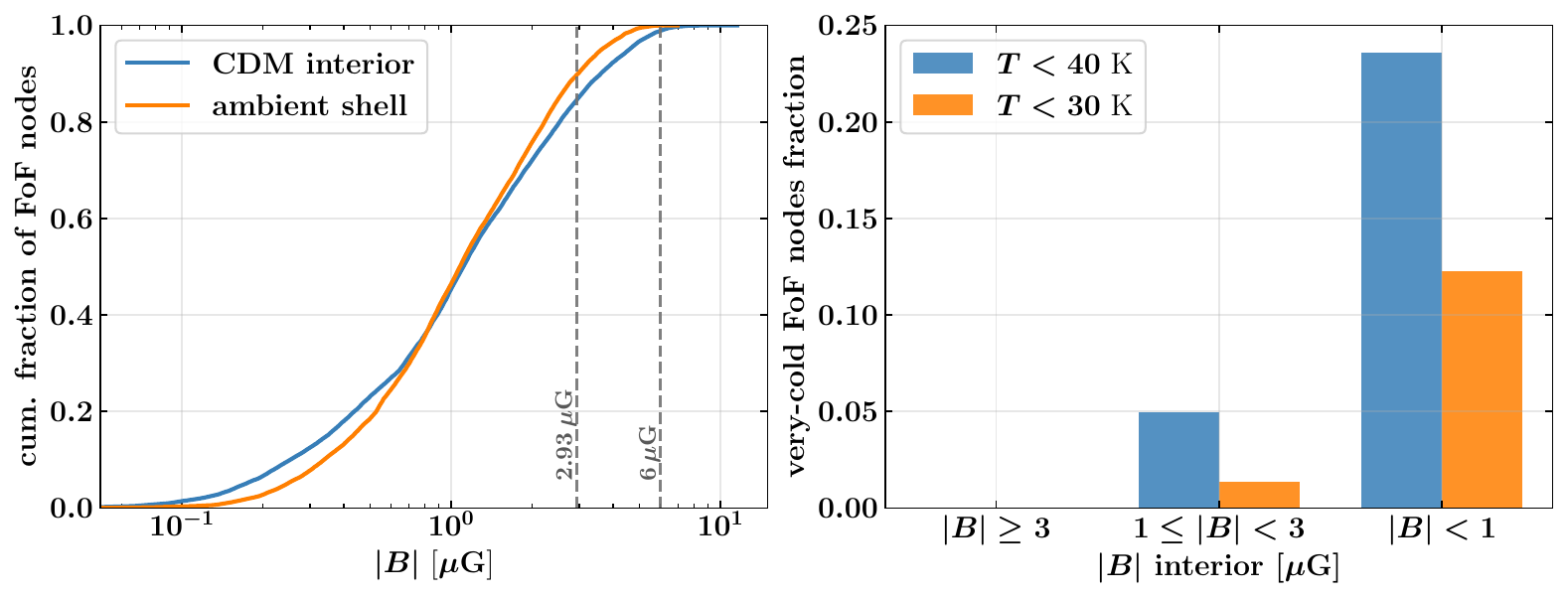}
  \caption{Sensitivity of the CDM properties to the evolved magnetic-field strength.
  The left panel shows the cumulative distributions of $|\mathbf{B}|$ inside the CDM and in the surrounding shell for the $t=100$--$200$~Myr sample, with vertical reference lines at $2.93$ and $6~\mu$G.
  The right panel shows the fraction of FoF nodes with median temperatures below $40$ and $30$~K in three bins of evolved interior magnetic-field strength.
  Although the simulation begins with $B_x=6~\mu$G, most CDM structures subsequently occupy substantially weaker fields, and the very cold population is concentrated almost entirely below $3~\mu$G.}
  \label{fig:mag_sensitivity}
\end{figure*}

We additionally tested whether the initial magnetic-field strength is important for the identified CDM by examining the evolved field strengths of the full FoF population.
Figure~\ref{fig:mag_sensitivity} shows the resulting field-strength distributions and their relation to the coldest CDM structures.
For the $t=100$--$200$~Myr sample, the interior field-strength distribution has $p_{16}/{\rm median}/p_{84}=0.36/1.11/2.88~\mu$G, while the surrounding shell has a median field strength of $1.08~\mu$G.
As shown in the left panel of Fig.~\ref{fig:mag_sensitivity}, 84.6~per cent of the CDM nodes have $B_\mathrm{in}\leq2.93~\mu$G, 85.2~per cent have $B_\mathrm{in}\leq3~\mu$G, and 98.9~per cent have $B_\mathrm{in}\leq6~\mu$G.

The coldest structures preferentially occupy still weaker magnetic fields, as illustrated in the right panel of Fig.~\ref{fig:mag_sensitivity}.
For nodes with median $T<40$~K, the median interior field strength is $0.40~\mu$G and 99.9~per cent lie below $2.93~\mu$G.
For $T<30$~K, the median decreases to $0.30~\mu$G and essentially all nodes lie below $3~\mu$G.
Among the persistent tracks, the correlation between track-median field strength and FoF lifetime is negligible ($\rho=0.037$), whereas the correlation between track-median field strength and minimum temperature is positive ($\rho=0.68$).
The simulation therefore provides no evidence that stronger evolved magnetic fields are required either to reach the lowest temperatures or to prolong the lifetime of connected CDM structures.

The pressure budget is consistent with this picture.
Within the same sample, the median interior pressure fractions are $f_\mathrm{CR}=0.963$, $f_\mathrm{mag}=0.035$, and $f_\mathrm{th}=4.3\times10^{-4}$, with a median ratio $P_\mathrm{mag}/P_\mathrm{CR}=0.037$.
Magnetic pressure is therefore important relative to the very small thermal pressure of the CDM, but it remains strongly subdominant to the CR pressure that establishes the dominant non-thermal pressure floor.

As an instantaneous post-processing sensitivity test, we rescaled the magnetic pressure according to $P_\mathrm{mag}\propto B^2$ while leaving the gas and CR state unchanged.
Reducing the field amplitude to $2.93/6$ of its original value increases the fraction of nodes within 0.1~dex of total inside--outside pressure balance from 97.1 to 99.7~per cent.
Even the formal $P_\mathrm{mag}=0$ limit leaves 99.6~per cent within 0.1~dex.
The boundary-normal pressure analysis shows the same qualitative trend, with magnetic pressure partly supporting the CDM interior such that weakening this component generally strengthens, rather than removes, the external compressive-pressure bias.

\subsection{Formation and assembly}

\begin{figure*}
  \centering
  \includegraphics[width=.99\linewidth]{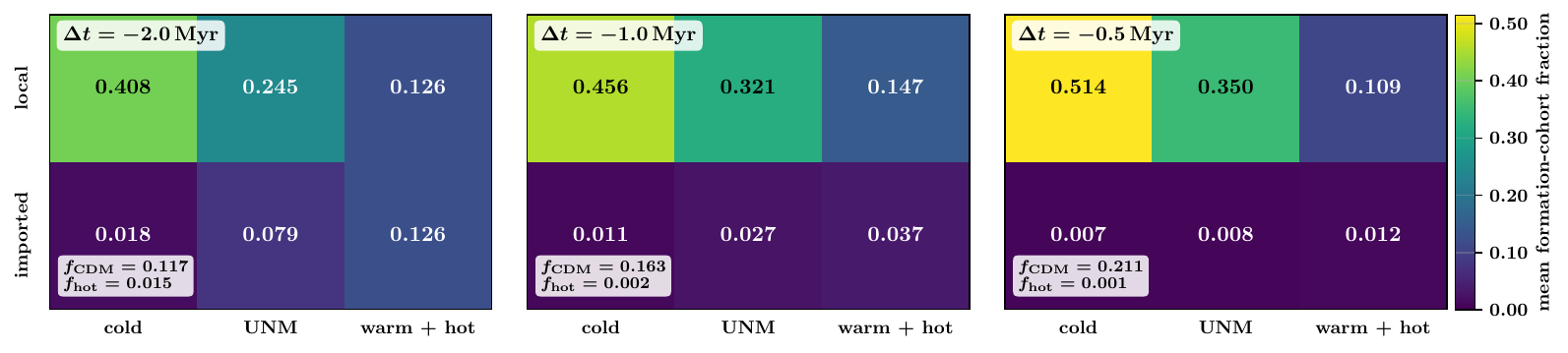}
  \caption{
  Mean origin fractions of the formation-cohort material at lookback times of $-2.0$, $-1.0$, and $-0.5$~Myr relative to the first detection of each CDM FoF structure.
  Rows distinguish material already local or imported from larger distances ($r>2\Delta x\approx7.8$~pc).
  Columns show material in the cold phase ($T<300$~K), the thermally unstable neutral medium (UNM; $300\leq T<5\times10^{3}$~K), and the combined warm and hot phases ($T\geq5\times10^{3}$~K).
  The inset values give the total fractions already satisfying the CDM definition, $f_\mathrm{CDM}$, and originating from the hot phase, $f_\mathrm{hot}$, with $T_\mathrm{hot}\geq3\times10^5$~K.
  }
  \label{fig:formation_assembly_matrix}
\end{figure*}

For each selected track, we defined the formation cohort as the tracers belonging to its first detected CDM FoF node and reconstructed their thermal state and location at fixed lookback times.
We classified a tracer as local when its minimum distance from the future first-node footprint was smaller than $2\Delta x\approx7.8$~pc and as imported otherwise.
The `imported' classification therefore indicates transport towards the future FoF structure, but does not imply that the material previously belonged to a separate cold cloud.
We distinguished pre-existing CDM, cold non-CDM gas with $T<300$~K, thermally unstable neutral-medium gas with $300\leq T<5\times10^{3}$~K, warm gas with $5\times10^{3}\leq T<3\times10^{5}$~K, and hot gas with $T\geq3\times10^{5}$~K.

Figure~\ref{fig:formation_assembly_matrix} shows that most formation-cohort material is already close to the future FoF footprint well before the structure is first detected.
The local fraction increases from approximately 0.78 at $-2.0$~Myr to 0.92 at $-1.0$~Myr and 0.97 at $-0.5$~Myr, while the imported fraction correspondingly decreases from 0.22 to 0.08 and 0.03.
This trend indicates that final assembly occurs predominantly close to the future CDM structure.
Crossing the adopted locality threshold of $2\Delta x\approx7.8$~pc over a 2~Myr lookback interval corresponds to a characteristic transport speed of $\sim3.8$~km~s$^{-1}$.

The thermal composition evolves more gradually.
The total cold fraction increases from approximately $f_\mathrm{cold}=0.43$ at $t_\mathrm{past}=-2.0$~Myr to $f_\mathrm{cold} = 0.47$ at $t_\mathrm{past}=-1.0$~Myr and $f_\mathrm{cold}=0.52$ at $t_\mathrm{past}=-0.5$~Myr, while the thermally unstable neutral medium (UNM) contributes approximately $f_\mathrm{UNM}=0.32$ to $0.36$ throughout.
The combined warm and hot contribution declines from approximately $f_\mathrm{hot}=0.25$ to $0.12$ over the same interval.
Only a minority of the cohort already satisfies the full CDM definition before first detection, and direct cooling from the hot phase contributes negligibly.
The formation cohorts therefore consist of a mixture of pre-existing cold material and gas cooling shortly before first detection, with the thermally unstable neutral medium providing a substantial fraction of the latter.
Such thermally unstable neutral gas is also observed in Galactic \ion{H}{1} absorption \citep[e.g.][]{Begum2010}.

The tracer histories thus favour local assembly from nearby cold and thermally unstable gas over the import of already-formed cold structures.
Direct hot-gas cooling is negligible in the present simulations.
However, in our chemical network, hot plasma is treated with equilibrium cooling rather than with a full time-dependent metal-ionisation network.
This may affect rapidly cooling supernova-heated gas \citep[e.g.][]{deAvillez2012,Breitschwerdt2021}.

For every FoF node, we additionally calculated the free-fall time from its mass density, $t_\mathrm{ff}=\sqrt{3\pi/(32G\rho)}$.
We find $t_\mathrm{ff} = 102^{325}_{14}$~Myr across the full population.
The CDM is therefore assembled and rearranged on timescales far shorter than its global free-fall time, favouring dynamical assembly over isolated gravitational collapse of the resolved structures.

\subsection{Cooling to very low temperatures}\label{sec:cooling}
To determine why the gas reaches temperatures of only a few tens of kelvin, we identified tracer transitions across $300\rightarrow100$~K, $100\rightarrow55$~K, and $55\rightarrow40$~K.
For each transition, we decomposed the temperature change into an ideal-gas adiabatic contribution, a contribution from changes in the mean molecular weight, and a residual non-adiabatic term, $\Delta\ln T_\mathrm{non-ad}=\Delta\ln T-(\gamma-1)\Delta\ln\rho-\Delta\ln\mu$.
A negative residual indicates net non-adiabatic cooling.
For a sample of 500 events per transition, we additionally extracted the specific thermal and CR energies, the unattenuated FUV field $G_0$, the local extinction $A_V$, and the magnetic-field strength.
We calculated the attenuated FUV field as $G_\mathrm{eff}=G_0\exp(-2.5A_V)$ and independently reconstructed the thermal, CR, and magnetic pressure components.
This post-processing diagnostic separates adiabatic from net non-adiabatic temperature changes and constrains the associated radiation and pressure environment, but it does not isolate individual microscopic heating and cooling channels.
The tracer thermodynamic histories show that adiabatic expansion is not the primary route by which gas enters the very cold CDM.
Of 4{,}398 identified $300\rightarrow100$~K cooling events, 96.3~per cent are dominated by the non-adiabatic term, whereas only 1.2~per cent are dominated by adiabatic expansion.
For the $100\rightarrow55$~K transition, which is most directly relevant to our CDM selection, 89.0~per cent of 4{,}763 events are dominated by non-adiabatic cooling and only 5.6~per cent by adiabatic expansion.
The median density ratio across this transition is $\rho_\mathrm{end}/\rho_\mathrm{start}=1.07$, and 57.5~per cent of the events are compressive rather than expansive.
The median adiabatic term therefore corresponds to slight compressional heating, so the inferred non-adiabatic cooling must both offset this heating and produce the observed temperature decrease.
Even among the 517 events cooling from 55 to 40~K, for which 87.2~per cent of the tracers are expanding, 80.7~per cent remain dominated by non-adiabatic cooling and the median adiabatic contribution accounts for only approximately 10~per cent of the total cooling.
Considering the ISRF, the temperature decrease is associated primarily with increasing attenuation rather than with a fading stellar radiation source.
During the $100\rightarrow55$~K transition, $G_\mathrm{eff}$ decreases in 64.2~per cent of the sampled events and $A_V$ increases in 62.2~per cent, whereas the unattenuated $G_0$ decreases in only 6.4~per cent.
Stronger non-adiabatic cooling is associated with larger decreases in $G_\mathrm{eff}$, with a Spearman coefficient of $\rho=-0.59$ when the cooling strength is quantified by $-\Delta\ln T_\mathrm{non-ad}$.
The corresponding relation with the approximate photoelectric-heating proxy is weaker, with $\rho=-0.32$.
This behaviour is consistent with the `variable-FUV' results of \citet{Rathjen2025}, where increased shielding together with the generally weak FUV field far from young clusters suppresses photoelectric heating and allows radiative and chemical cooling to lower the gas temperature to a few tens of kelvin.

Thermal conduction cannot be responsible for producing the simulated cold phase because it is not included in the present simulations.
However, unresolved conductive and turbulent-mixing interfaces remain an important caveat when extrapolating the simulated survival times to real clouds.

For the $100\rightarrow55$~K events, the CR pressure fraction increases in 51.8~per cent of the 500-event diagnostic sample.
This increase correlates only modestly with the adiabatic contribution ($\rho=0.27$) and anticorrelates weakly with the density change ($\rho=-0.28$).
Although these trends are consistent with some CR-associated expansion, this effect is secondary because only 5.6~per cent of the events are dominated by adiabatic expansion.
We therefore interpret the primary role of CRs as dynamical support, as they provide a pervasive non-thermal pressure component that allows gas with very low thermal pressure to remain diffuse after cooling rather than directly driving the temperature decrease.

\subsection{Lifetimes}

\begin{figure}
  \centering
  \includegraphics[width=.98\linewidth]{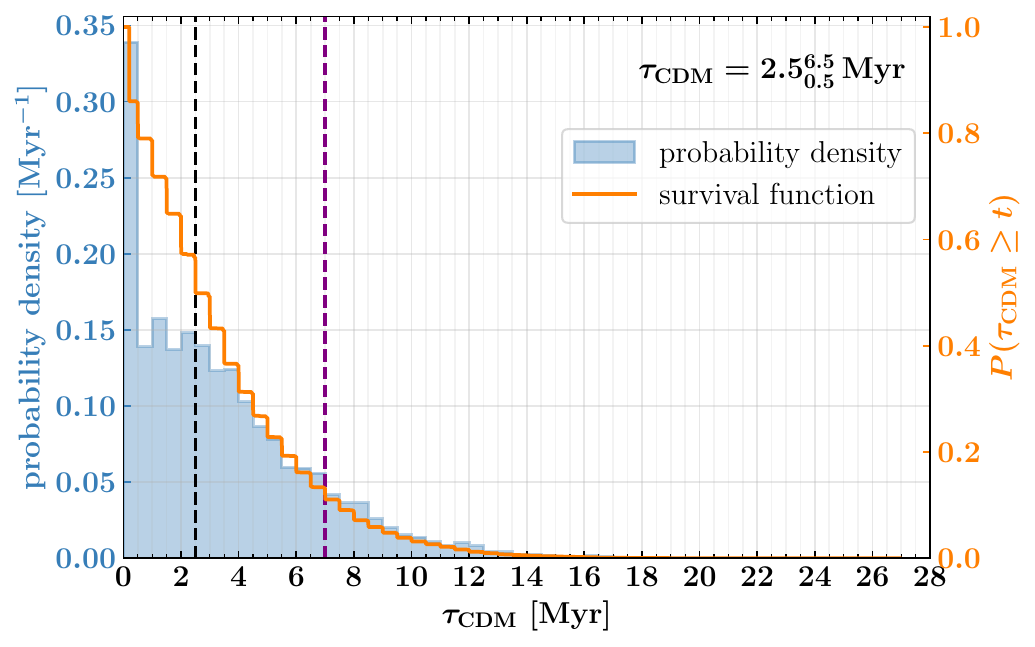}
  \caption{
  Probability density and empirical survival function of the CDM lifetimes, $\tau_\mathrm{CDM}$.
  The blue histogram corresponds to the left-hand axis, while the orange empirical survival function, $P(\tau_\mathrm{CDM}\geq t)$, corresponds to the right-hand axis.
  The sample combines the 150 persistent tracks with the highest peak masses and the 150 tracks with the lowest median temperatures, yielding 300 selected tracks and 46{,}933 unique sampled tracer particles.
  We sampled up to 250 tracers per track and followed them at a median cadence of 0.5~Myr.
  The black dashed vertical line marks the median CDM lifetime, while the purple dashed vertical line marks the proposed 7~Myr interval since the CLIC, and potentially the LLCC, entered the LB.
  We find $\tau_\mathrm{CDM}=2.5^{6.5}_{0.5}$~Myr, where the bounds give the 16th and 84th percentiles.
  }
  \label{fig:sampled_tracer_residence}
\end{figure}

We defined the CDM lifetime, $\tau_\mathrm{CDM}$, as the longest contiguous interval during which an individual tracer associated with a specific FoF structure continuously satisfies the CDM selection criteria.
Figure~\ref{fig:sampled_tracer_residence} shows the probability density and empirical survival function of $\tau_\mathrm{CDM}$ for the sampled peak-cohort tracer particles drawn from the 300 selected tracks.
The distribution is broad, with $\tau_\mathrm{CDM}=2.5^{6.5}_{0.5}$~Myr.
The quoted bounds are the 16th and 84th percentiles of the sampled distribution.
The empirical survival function gives $P(\tau_\mathrm{CDM}\geq7~\mathrm{Myr})=0.125$ for the sampled population.

For numerical reasons, we sampled the tracers at a cadence of 0.5~Myr, so their histories are coarse-grained on this timescale.
Furthermore, the sample deliberately favours massive and particularly cold persistent tracks.
The measured value should therefore be interpreted as the characteristic lifetime of the sampled peak-cohort material rather than as an unbiased estimate for all CDM material in the simulation.

\subsection{Fate of the CDM}

\begin{figure}
  \centering
  \includegraphics[width=.98\linewidth]{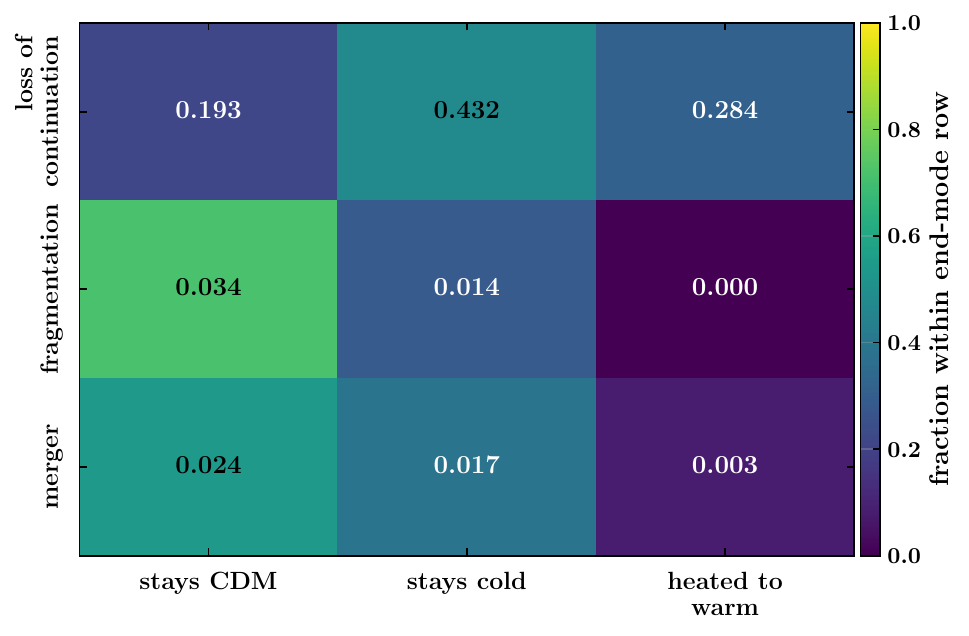}
  \caption{
  Thermal fate of the tracer cohorts after the end of their associated FoF identities, separated by topological ending mode.
  The colour bar indicates the fraction within each ending mode that reaches the corresponding thermal state, while the number in each tile gives the total fraction of selected tracks with that combination of thermal state and ending mode.
  Most cohorts remain either CDM or in the broader $T<300$~K cold phase after the FoF identity ends.
  No cohort is predominantly heated into the hot phase.
  }
  \label{fig:track_endings_matrix}
\end{figure}

Figure~\ref{fig:track_endings_matrix} cross-classifies the topological ending modes with the thermal fate of the associated tracer cohort at the first available post-ending state.
Of the 300 selected tracks, 296 have an observed ending, and four are right-censored, i.e. their tracks have not terminated by the end of the simulated evolution.

Of the observed endings, 269 result from loss of a significant continuation, 14 from fragmentation into smaller identified CDM structures, and 13 from mergers with other structures.
Loss of continuation can occur when gas dynamics or feedback disperses the FoF structure, or when the tracked material is heated sufficiently to leave the CDM phase.

At the first post-ending state, 74 of the 296 cohorts remain in the CDM phase, 137 remain below 300~K but no longer satisfy the CDM definition, and 85 are predominantly heated into the warm phase.
Thus, 211 cohorts, or 71.3~per cent, remain predominantly cold after the FoF identity ends.
No selected cohort is predominantly heated into the hot phase, as would be expected if engulfment by supernova remnants were a common destruction mechanism.
However, nearby supernova feedback may still shear, accelerate, or fragment cold gas without heating most of its mass above the adopted threshold.

Among tracks ending through loss of a significant continuation, approximately 69~per cent remain in the cold phase.
The disappearance of a FoF identity therefore usually marks a loss of spatial coherence rather than thermochemical destruction of the cold material.
Such reconfiguration may result from turbulent shear, ram-pressure stripping, feedback-driven acceleration, or other dynamical processes.

\subsection{The role of nearby supernovae}\label{sec:sne}
The analysed simulation interval contains 753 core-collapse supernova events.
For each observed CDM track endpoint, we determined the distance to the closest preceding supernova within a 1~Myr lookback window and counted events within 25, 50, and 100~pc.

None of the 85 cohorts heated into the warm phase in the full sample has a preceding supernova within 50~pc.
Only five of the heated-to-warm endpoints have a preceding event within 100~pc.
By comparison, 28 of the cold gas survivors have a preceding event within 100~pc.
The fractions with an event within 100~pc are 5.9~per cent for heated-to-warm episodes and 15.0~per cent for cold survivors.
These statistics do not support a scenario in which an individually identifiable, recent nearby supernova is the dominant immediate cause of heating or identity loss.

Although most selected endpoints cannot be attributed directly to a recent nearby supernova, supernovae remain essential for generating the hot, turbulent, magnetised, and CR-supported environment.
They can also shear, accelerate, fragment, or disconnect cold structures without heating most of their mass.

\section{Discussion and conclusions}\label{sec:discussion}

The LLCC raises two central questions concerning whether cold diffuse gas can form inside the LB and whether it can subsequently survive there for several Myr.
The \textsc{SILCC} simulations provide affirmative answers to both questions in a generic solar-neighbourhood, feedback-regulated environment without requiring either direct condensation from hot plasma or the import of an already-formed cold cloud.
They do not reconstruct the specific history of the LLCC, but establish a physically viable pathway for producing and maintaining LLCC-like cold material.

The simulated CDM is predominantly filamentary or sheet-like and lies preferentially far from young stellar clusters, where a weaker ISRF reduces photoelectric heating and molecular photodissociation.
Warm gas constitutes its characteristic immediate environment, while substantial hot-gas contact occurs for only approximately 5.0~per cent of the population.
The CDM should therefore be regarded as cold gas embedded in a turbulent, multiphase warm--hot medium rather than as isolated clouds uniformly immersed in million-degree plasma.

The strong thermal-pressure contrast across the CDM interfaces is largely compensated by non-thermal pressure.
The CR pressure remains nearly continuous between the cold and ambient gas, while magnetic pressure provides an additional contribution, bringing the summed thermal, CR, and magnetic pressures close to balance.
Cosmic rays therefore act primarily as a broadly distributed pressure floor rather than as a local pressure enhancement around individual structures.

This interpretation is not sensitive to the initially adopted magnetic-field strength.
Although the simulation starts with $B_x=6~\mu$G, the evolved CDM has a median field strength of only $1.11~\mu$G, and the coldest structures occur almost exclusively below $3~\mu$G.
Magnetic pressure contributes only a few per cent of the median summed pressure, and reducing this component in post-processing leaves the approximate pressure balance essentially unchanged.
These evolved field strengths are also consistent with local observational constraints of a few $\mu$G \citep{Zirnstein2016,Hutschenreuter2024}.

\paragraph{Could the LLCC have formed inside the Local Bubble?}

At $t_\mathrm{past}=-0.5$~Myr, approximately 52~per cent of the formation-cohort material is already below 300~K, 36~per cent lies in the thermally unstable neutral medium, and only 12~per cent is warm or hot.
The material also becomes increasingly concentrated near the future FoF footprint as first detection approaches.
The simulations therefore favour local assembly from a mixture of pre-existing cold material and nearby thermally unstable gas that cools shortly before the structure is identified.
Import of already-formed cold fragments and direct condensation from hot gas are both minor channels.

\paragraph{Why is the CDM very cold?}

The tracer-particle analysis shows that the transition from the thermally unstable regime to the CDM is dominated by non-adiabatic cooling.
This applies both from 300 to 100~K and at lower temperatures, where 89.0~per cent of the $100\rightarrow55$~K transitions and 80.7~per cent of the $55\rightarrow40$~K transitions are dominated by the non-adiabatic term.
The $100\rightarrow55$~K transition is also more frequently compressive than expansive, arguing against adiabatic expansion as the general origin of the lowest temperatures.
Cooling is instead associated with increasing attenuation and decreasing $G_\mathrm{eff}$, while the unattenuated $G_0$ generally remains unchanged.
Enhanced FUV shielding therefore suppresses photoelectric heating and allows radiative and chemical cooling to continue from the thermally unstable neutral medium to temperatures of 20--50~K.
Thermal conduction is not included, while CR-associated expansion contributes only secondarily.
The principal role of CRs is mechanical, providing non-thermal support that allows the resulting low-thermal-pressure gas to remain diffuse after cooling.

\paragraph{Could the LLCC survive for 7~Myr inside the Local Bubble?}

The lifetime analysis yields $\tau_\mathrm{CDM}=2.5^{6.5}_{0.5}$~Myr, and 12.5~per cent of the sampled peak-cohort tracers remain continuously in the CDM for at least 7~Myr.
The 7~Myr interval is motivated by the proposed time since material associated with the CLIC entered the LB.
Cold diffuse material can therefore survive continuously for several Myr in the simulated feedback-driven environment.

The free-fall time of the resolved CDM structures is approximately 100~Myr, with a median lifetime-to-free-fall-time ratio of only $\sim10^{-2}$.
Their evolution is therefore much faster than global gravitational collapse.
Furthermore, 71.3~per cent of tracer cohorts remain predominantly cold after their associated FoF identity disappears, showing that loss of spatial continuity usually represents dynamical rearrangement rather than thermochemical destruction.

Recent nearby supernovae are not the dominant immediate destruction mechanism in the selected sample.
Neither warm-heated nor hot-contact endpoints occur within 50~pc of a preceding supernova during the adopted 1~Myr lookback interval.
Supernovae nevertheless remain important indirectly by generating the hot gas, turbulence, and CR population that shape the environment in which the CDM evolves.

\paragraph{Synthesis}

The \textsc{SILCC} results provide a turbulent, multiphase counterpart to earlier idealised converging-flow models \citep[e.g.][]{Audit2005,Heitsch2006,VazquezSemadeni2006}.
As in these models, cold-gas formation involves compression, thermal instability, radiative cooling, and the formation of filamentary or sheet-like cold structures.
However, we do not find a single coherent collision between warm clouds as the characteristic formation mechanism.
Instead, CDM structures assemble from spatially and thermally heterogeneous material within a turbulent multiphase medium.

The CDM-interface analysis supports this distinction.
The pressure field shows a mild statistical bias towards compression, which is stronger when persistent structures are first detected, but $\partial v_n/\partial n$ does not reveal a corresponding population-wide converging flow.
Only the subset with substantial hot-gas contact shows a clearer compression signature.
Local dynamical compression therefore occurs, but is not a universal formation mechanism.

The pressure structure also differs qualitatively from purely hydrodynamic converging-flow models.
Whereas \citet{VazquezSemadeni2006} found the cold phase to be thermally overpressured relative to its warm surroundings, the simulated CDM is generally thermally underpressured.
Approximate pressure balance is recovered only after including the non-thermal CR and magnetic components.

The otherwise identical simulation without CR transport further shows that the cold diffuse branch is strongly suppressed when CRs are omitted.
Combined with the cooling analysis, this points to a primarily mechanical role for CRs.
They do not directly drive the temperature decrease, but provide a broadly distributed non-thermal pressure reservoir that allows gas to remain diffuse after losing thermal energy.
Without this pressure floor, maintaining low-density gas at temperatures of only a few tens of kelvin becomes more difficult, favouring compression or rearrangement towards denser material.

Taken together, the simulations establish a physically viable pathway in which cold gas assembles locally, cools to a few tens of kelvin, and survives for several Myr within a warm--hot superbubble environment.
They do not determine the actual formation history of the LLCC, but demonstrate that its existence inside the LB does not require it to have entered as an already-formed cold cloud.

\subsection{Caveats}\label{sec:caveats}

The \textsc{SILCC} simulations do not model the LB or the LLCC directly.
Their feedback environment is more sustained and energetic than that inferred for the present-day LB, and the simulated CDM structures are not one-to-one analogues of the observed LLCC in density, column density, or thickness.
Most importantly, the spatial resolution of $\Delta x\approx3.9$~pc cannot resolve the observed 0.25--0.54~pc width of the LLCC, its internal substructure, or its conductive and turbulent-mixing interfaces.
The simulations should therefore be interpreted as testing whether cold gas can form and survive under self-consistent solar-neighbourhood feedback conditions rather than as a direct reconstruction of the LLCC.

The adopted magnetic-field strength is another model assumption.
The evolved CDM is predominantly found at $B\lesssim3~\mu$G, and its instantaneous pressure balance is insensitive to a substantial reduction of the magnetic-pressure term.
However, the periodic $500\times500~\mathrm{pc}^2$ midplane domain does not include differential galactic rotation or large-scale shear, while unresolved small-scale dynamo amplification is also absent at $\Delta x\approx3.9$~pc.
The magnetic field may therefore decline through numerical diffusion without being replenished by these processes \citep[see e.g.][and references therein]{Beck2019}.
Earlier \textsc{SILCC} studies indicate that magnetic fields primarily retard gravitational collapse.
At the same time, the adopted anisotropic CR diffusion depends mainly on the field direction rather than directly on its strength \citep{Walch2015,Girichidis2018}.

The tracer sample is intentionally biased towards persistent structures with high peak masses or low median temperatures and is therefore not representative of the full CDM population.
The 0.5~Myr sampling cadence can also miss shorter excursions from the selected phase, thereby coarse-graining the inferred residence times.

Finally, the reconstruction of formation histories and FoF identities depends on the adopted $2\Delta x$ local--imported threshold, connectivity, and overlap criteria.
These choices affect detailed classifications and absolute numbers, although the independent Eulerian tracking test indicates that the overall lifetime distribution is robust.

\begin{acknowledgements}
Funded by the Deutsche Forschungsgemeinschaft (DFG, German Research Foundation) -- Project-ID 500700252 -- SFB 1601.

Funded by the Deutsche Forschungsgemeinschaft (DFG, German Research Foundation) under Germany's Excellence Strategy EXC 3037 -- 533607693 -- Unser dynamisches Universum.

J.L.L. acknowledges support from the University of Colorado Foundation.

The authors thank the Space Telescope Science Institute (STScI) for hosting the 2025 STScI Spring Symposium, during which the idea for this work originated.

This work made use of FLASH \citep{Fryxell2000,Dubey2009}, yt \citep{Turk2011}, NumPy \citep{vanderWalt2011}, Matplotlib \citep{Hunter2007}, h5py \citep{Collette2020}, and IPython \citep{Perez2007}.
\end{acknowledgements}

\section*{Data availability}
The derived data underlying this article are available from the corresponding author on reasonable request.
The simulation data are available from the \href{http://silcc.mpa-garching.mpg.de/}{\textsc{SILCC} Project data page}.

\bibliographystyle{aa}
\bibliography{lit}

\begin{appendix}
\section{Summary properties of the CDM population}\label{app:population_summary}
\begin{table}[!ht]
  \centering
  \caption{
  Summary of the principal properties of the identified CDM population.
  Unless stated otherwise, values refer to all 47{,}045 FoF nodes and give the median, with the 16th and 84th percentiles where available.
  Interface diagnostics are quoted at a separation of three grid cells ($3\Delta x\approx11.7$~pc).
  }
  \setlength{\tabcolsep}{6pt}
  \begin{tabular}{lll}
  \hline
  Property & Symbol & Value \\
  \hline
  \multicolumn{3}{l}{\textit{Structure and environment}} \\
  Number of FoF nodes
  & $N_\mathrm{FoF}$
  & 47{,}045 \\
  Mass
  & $M$
  & $84.7^{6500}_{8.8}\,M_\odot$ \\
  Number of cells
  & $N_\mathrm{cell}$
  & 219 \\
  Equivalent radius
  & $R_\mathrm{eq}$
  & $14.6^{19.9}_{12.7}$~pc \\
  Internal temperature
  & $T_\mathrm{in}$
  & $50.9^{53.3}_{44.0}$~K \\
  Ambient temperature
  & $T_\mathrm{out}$
  & $1.16^{4.30}_{0.29}\times10^{3}$~K \\
  Partially warm-embedded fraction
  & $0.1\leq f_\mathrm{warm,out}<0.5$
  & 18.87~per cent \\
  Strongly warm-embedded fraction
  & $f_\mathrm{warm,out}\geq0.5$
  & 81.11~per cent \\
  Hot-gas contact
  & $f_\mathrm{hot,out}\geq0.1$
  & 5.01~per cent \\
  Distance to nearest active OB cluster
  & $d_\mathrm{OB}$
  & $159^{326}_{73}$~pc \\
  Distance to nearest \ion{H}{2}-region boundary
  & $d_{\mathrm{H\,II}}$
  & $14^{55}_{6}$~pc \\
  \hline
  \multicolumn{3}{l}{\textit{Internal and ambient pressure properties}} \\
  Internal thermal pressure
  & $P_\mathrm{th,in}/k_\mathrm{B}$
  & $7.7$~K~cm$^{-3}$ \\
  Ambient thermal pressure
  & $P_\mathrm{th,out}/k_\mathrm{B}$
  & $262$~K~cm$^{-3}$ \\
  Internal magnetic-field strength
  & $|\mathbf{B}|_\mathrm{in}$
  & $1.44~\mu$G \\
  Ambient magnetic-field strength
  & $|\mathbf{B}|_\mathrm{out}$
  & $1.39~\mu$G \\
  Internal CR pressure
  & $P_\mathrm{CR,in}/k_\mathrm{B}$
  & $9.12\times10^{3}$~K~cm$^{-3}$ \\
  Ambient CR pressure
  & $P_\mathrm{CR,out}/k_\mathrm{B}$
  & $9.14\times10^{3}$~K~cm$^{-3}$ \\
  Internal summed pressure
  & $P_\mathrm{sum,in}/k_\mathrm{B}$
  & $1.04\times10^{4}$~K~cm$^{-3}$ \\
  Ambient summed pressure
  & $P_\mathrm{sum,out}/k_\mathrm{B}$
  & $1.06\times10^{4}$~K~cm$^{-3}$ \\
  \hline
  \multicolumn{3}{l}{\textit{Three-cell interface diagnostics}} \\
  Summed-pressure contrast
  & $\Delta P_\mathrm{sum}/k_\mathrm{B}$
  & $+50^{+342}_{-342}$~K~cm$^{-3}$ \\
  Thermal-pressure contrast
  & $\Delta P_\mathrm{th}/k_\mathrm{B}$
  & $+95$~K~cm$^{-3}$ \\
  CR-pressure contrast
  & $\Delta P_\mathrm{CR}/k_\mathrm{B}$
  & $+3$~K~cm$^{-3}$ \\
  Magnetic-pressure contrast
  & $\Delta P_\mathrm{mag}/k_\mathrm{B}$
  & $-38$~K~cm$^{-3}$ \\
  Effective compression pressure
  & $P_\mathrm{comp,eff}/k_\mathrm{B}$
  & $+161$~K~cm$^{-3}$ \\
  Compressing boundary fraction
  & $f(P_\mathrm{comp,eff}>0)$
  & 0.65 \\
  Boundary-normal velocity gradient
  & $\partial v_n/\partial n$
  & $-0.0009$~Myr$^{-1}$ \\
  Converging boundary fraction
  & $f(\partial v_n/\partial n<0)$
  & 0.50 \\
  \hline
  \end{tabular}
  \tablefoot{
  Positive pressure contrasts indicate larger pressure outside the CDM structure.
  Negative boundary-normal velocity gradients indicate converging flow.}
  \label{tab:cdm_population_properties}
\end{table}

\end{appendix}

\end{document}